\documentclass{cta-author}

{}
{}
\newtheorem{remark}{Remark}{}

\usepackage{url}
\usepackage{amssymb}
\usepackage{amsmath}
\usepackage{booktabs} 
\usepackage{tabularx} 
\usepackage{subfigure}
\usepackage{algorithmic}
\usepackage{graphics}
\usepackage{appendix}
\usepackage{textcomp}
\usepackage{subfigure}
\usepackage{tabularx}
\usepackage{times}
\usepackage{multirow}
\usepackage{bm}
\usepackage{latexsym}
\usepackage{booktabs}
\usepackage{threeparttable}
\usepackage{epsf}
\usepackage{algorithm}
\usepackage{color,colortbl}

\usepackage{multicol}
\usepackage{graphicx}
\usepackage{epstopdf}
\usepackage{multirow}
\usepackage{threeparttable} 
\usepackage{diagbox} 
\usepackage{array}
\usepackage{natbib}

\begin{document}

\supertitle{Brief Paper}

\title{Ambiguity Function Shaping based on Alternating Direction Riemannian Optimal Algorithm}

\author{\au{Haoyu Yi} \au{Xinyu Zhang$^*$} \au{Weidong Jiang} \au{Kai Huo}}

\address{{College of Electronic Science and Technology, National University of  Defense Technology, Changsha, 410073, People’s Republic of China}
\email{zhangxinyu90111@163.com}}

\begin{abstract}
\looseness=-1  
In order to improve the ability of cognitive radar (CR) to adapt to the environment, the required ambiguity function (AF) can be synthesized by designing the waveform. 
The key to this problem is how to minimize the interference power. 
Suppressing the interference power is equivalent to minimize the expectation of slow-time ambiguity function (STAF) over range-Doppler bins.
From a technical point of view, this is actually an optimization problem of a non-convex quartic function with constant modulus constraints (CMC). 
In this paper, we proposed a novel method to design a waveform to synthesize the STAF based on suppressing the interference power. 
We put forward alternating direction Riemannian optimal algorithm within an alternating direction penalty method (ADPM) framework. 
In each iteration, this problem is split into two sub-problems by introducing auxiliary variables. In the first sub-problem, we solved the convex problem directly with a closed-form solution, then utilized the Riemannian trust region (RTR) algorithm in the second sub-problem. Simulation results demonstrate that the proposed algorithm outperforms other advanced algorithms in the aspects of STAF, range-cut and signal-to-interference-ratio (SIR) value.
\end{abstract}

\maketitle

\section{Introduction}\label{sec1}

The concept of cognitive radar (CR) was first introduced by Simon Haykin \cite{1}. In a CR system, the receiver, transmitter, and environment form a dynamic closed loop. Unlike traditional radar, the received waveform containing environmental knowledge can be extracted to feed back to the transmitter \cite{2}. The ambiguity function (AF) in a radar system is a common tool used to evaluate the performance of radar waveform, reflecting the range-Doppler response at the output of the matched filter \cite{3}. Utilizing prior information provided by CR, the method of shaping a reasonable AF is to minimize the range-Doppler bin of interest. Several studies have been proposed for different scenarios \cite{4}\cite{5}. 
In slow-time coding, the range-Doppler response at the output of the matched filter can be considered as a small "slice" of the AF at zero-time delay and multiples of pulse repetition interval (PRI). This response is called the slow-time ambiguity function (STAF) \cite{4}. In a practical radar system, the received radar echoes include both target echoes and echoes from interfering scatterers \cite{5}. This mechanism relies on minimizing (suppressing) the average value of the STAF over the scatterer position. Thus, that is corresponding to minimize the interference power at the output of the matched filter.

We consider the shaping of the range-Doppler response of the slow-time transmitted waveform of a CR system. Radar systems can dynamically predict the scattering environment by utilizing information such as meteorological data, terrain information, and previous electromagnetic scatterer distribution \cite{6}\cite{7}. Therefore, our intention is to suppress the interference power at the output of the matched filter using a prior knowledge. Due to the presence of nonlinear power amplifiers \cite{8}-\cite{10}, the signal must operate in saturation to maximize efficiency. Coupled with the cost constraint of the transmitted signal, a constant modulus constraint (CMC) on the signal is necessary.

Therefore, under the constant modulus constraint, the optimization problem of STAF shaping is formulated as a complex quartic non-convex polynomial problem \cite{11}-\cite{14}. To address this challenge, the shaping of non-periodic AF of unimodular sequence was tackled by accelerating iterative sequential optimization (AISO) algorithm \cite{15}, which actually belongs to the majorization-minimization framework \cite{16}. In \cite{17}, the authors employed an optimization procedure based on the maximum block improvement (MBI) method. In addition, the CMC is dealt with by generating a large number of random samples, combined with the MBI algorithm to achieve a more effective solution. Wu \textit{et al.} developed a method by using majorization-minimization (MM) and the coordinate descent method \cite{18}. This method focuses on transforming the quadratic function into quadratic form, followed by the coordinate descent method. However, that is not effective enough for direct optimization of non-convex constraints. In recent work, Cui \textit{et al} proposed an inexact alternating direction penalty method (ADPM) framework \cite{19} with iterative sequential quartic optimization (ISQO) algorithm \cite{20}. This algorithm achieves better performance by relaxing condition of the cost function, but required much more computational time. Khaled \cite{5} employed the principle of optimization on non-convex Riemannian manifolds to solve this problem. In his work, a Riemann quartic descent algorithm was developed which can descend on complex circular manifolds, but convergence is slightly slower and requires a large number of iterations as only first-order information is used. In \cite{21}, Qiu \textit{et al.} derived the Riemann Hessian matrix of the cost function and developed a second-order gradient-based quadratic Riemannian trust region (RTR) algorithm with a satisfied result. However, their methods only converge to a local optimum and there is still a space for further optimization.

In this paper, we developed a new method which can further optimize the original solution. Considering the CMC, we developed an efficient iterative algorithm based on the ADPM framework \cite{22}-\cite{25} and the RTR algorithm, which provides a better solution. Specifically, utilizing RTR algorithm can directly update the transmitted waveform in each iteration, which involves to optimizing a quartic problem under CMC. To this end, we introduce an efficient ADPM framework to implement parallelly in each iteration in order to obtain a closed-form solution. The main contributions include the following.
\begin{enumerate}[(iii)]
\item[(i)]We utilize an efficient ADPM framework. This framework for non-convex problems asymptotically converges to a primal feasible point under mild conditions \cite{22}. Due to the non-convexity of the cost function, we introduced the RTR algorithm for the second iteration within the ADPM framework.
\item[(ii)]We derive RTR algorithm in ADPM framework. RTR algorithm can escape from latent saddle points and achieve a super-linear convergence rate. We derive the Riemannian first-order gradient and the Hessian matrix of the cost function with penalty terms in the second sub-problem of the framework to ensure convergence of the iterations.
\item[(iii)]We compare the performance among three algorithms and evaluate the effect of the optimization. One is the inexact ADPM+ISQO algorithm and only RTR algorithm is the other one. Compared with inexact ADPM+ISQO algorithm , our proposed method spends less time to reach the terminal condition and achieves a better effect. On the other hand, relative to only RTR algorithm, the cost function can be optimized by our proposed algorithm despite sacrificing more time to iterate. The further suppression interference effect can be reflected by range-cut along STAFs. 
\end{enumerate}
The rest of this paper is organized as follows. The signal model is presented in Section II. In Section III, we solve the two sub-problems separately. Section IV compares the proposed algorithm with state-of-the-art methods and section V concludes the paper.
Notations: We use boldface upper case for matrices and boldface lower case for vectors. The specific notations are all listed in Table 1. 
\begin{table}[!h]
\processtable{List of notations in this paper\label{tab1}}
{\begin{tabular*}{20pc}{@{\extracolsep{\fill}}lll@{}}\toprule
$\textbf{Notation}$ & $\textbf{Meanings}$ \\\midrule
${\{}\cdot{\}^{T}}$ & Transpose \\ 
${\{}\cdot{\}^{H}}$ & Conjugate transpose  \\ 
$\mathbf{I}_K$ & Identity matrix of size K × K \\ 
diag$(\mathbf{x})$ & Diagonal matrix comprised of the vector $\mathbf{x}$ \\
$\{\cdot\}^*$ & Conjugate\\
$\odot$ & Hadamad product \\
$\mathbb{C}$ & Complex number set\\
$\mathbb{R}$ & Real number set\\
$\mathbb{N}$ & Integer set\\
$\mathbb{C}^{K}$ & K-dimensional complex-valued vectors\\
$\mathbb{R}^{K}$ & K-dimensional real-valued vectors\\
$\mathbb{E}$ & Statistical expectation\\
$\vert\mathbf{\cdot}\vert$ & Modulus\\
$\Vert \mathbf{\cdot} \Vert$ & $l_2$ norm\\
$\mathcal {R}\{\cdot\}$ & Real part\\
$\mathcal {I}\{\cdot\}$ & Real part\\
$\langle\cdot\rangle$ & Inner product \\\bottomrule
\end{tabular*}}{}
\end{table}

\section{Signal Model}\label{sec2}
Focusing on CR system, consider a slow-time radar code $\mathbf{s}=[\mathbf{s}(0),\mathbf{s}(1),..,\mathbf{s}(K-1)]^{T}\in\mathbb{C}^{K}$. The waveform from the receiver is down-converted at the base band and passes through a pulse matched filter, then it is sampled into digital signal. The observed K-dimensional vector $\mathbf{z}=[\mathbf{z}(0),\mathbf{z}(1),..,\mathbf{z}(K-1)]^{T}\in\mathbb{C}^{K}$ from the range-azimuth cell can be formulated as
\begin{align}\label{eq1}
\mathbf{z}=\alpha\mathbf{y}(\mathbf{s},v_{d_T})+\mathbf{c}+\mathbf{n}
\end{align}
where $\alpha$ is a complex parameter including back scattering effects and channel propagation within the range-azimuth bin of interest. Vector $\mathbf{n}$ represents received additive noise, with 
$\mathbb{E}[\mathbf{n}]=0$ and $\mathbb{E}[\mathbf{nn^{H}}]=\sigma^{2}_n\mathbf{I}$. Vector $\mathbf{c}$ denotes the clutter vector located at different range-azimuth bins $(r,l)$, where $r \in \{0,1,...,K-1\}$, $l \in \{0,1,...,L\}$ (as illustrated in Fig. 1).
\begin{figure}[htpb]
\centering
\includegraphics[width=3in]{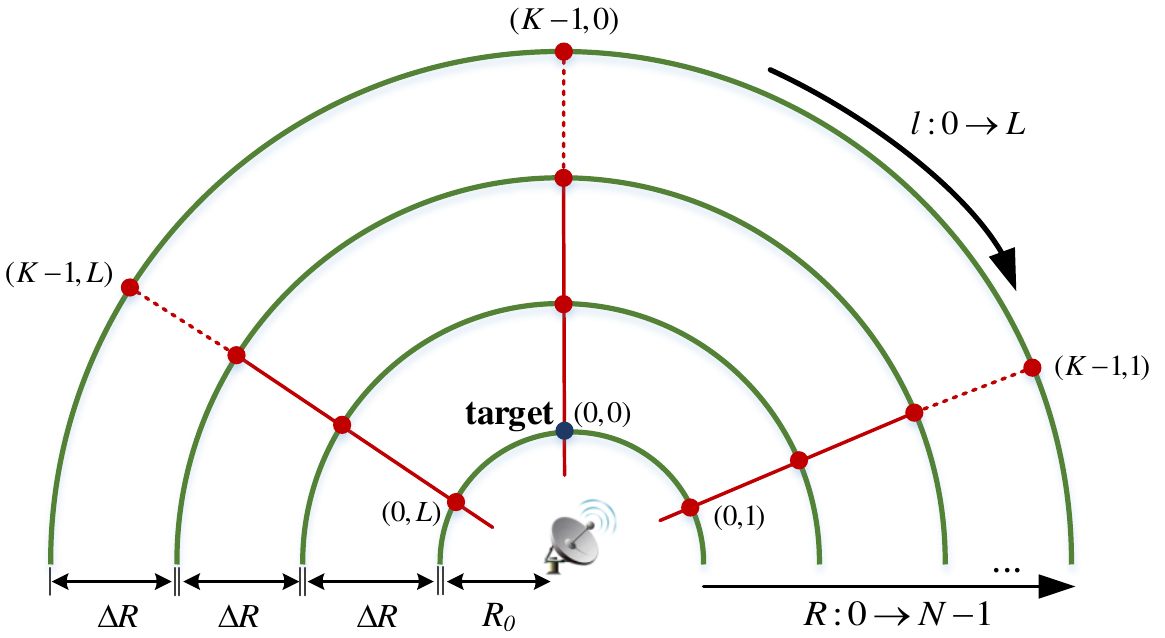}
\caption{Range-azimuth bins.}
\end{figure}
The specific equation of $\mathbf{y}$ is 
\begin{align}\label{eq2}
\mathbf{y}(\mathbf{s},v_{d_T})=\mathbf{s}\odot\mathbf{p}(v_{d_T})
\end{align}
where $\mathbf{p}(v_{d_T})=[1,e^{2\pi{v_{d_T}}},...,e^{2\pi{(K-1)v_{d_T}}}]\in\mathbb{C}^{K}$ is the temporal steering vector with $v_{d_T}$ the normalized target Doppler frequency. Vector $\mathbf{c}$ in (1) can be expressed as 
\begin{align}\label{eq3}
\mathbf{c}=\sum_{i=1}^{N_t}\gamma_i\mathbf{J}_{r_i}\mathbf{y}(\mathbf{s},v_{d_i})
\end{align}
where ${N_t}$ is the total number of interference scatterers, $k\in\{0,1,...,K-1\}$ denotes the range position, $\gamma_i$ is the complex echo amplitude, and $v_{d_i}$ represents the normalized Doppler frequency of the $i$th interference scatter. Considering the range ambiguity effect in the radar system, scattering points in adjacent range positions may interfere the target echo. The shift matrix $\mathbf{J}_r$ is given as
\begin{equation}\label{eq4}
\mathbf{J}_r(m,n)=\left\{
\begin{array}{rcl}
1& & {if\quad m-n=r}\\
0& & {if\quad m-n\neq{r}}\\
\end{array} \right. 
\end{equation}
with $\mathbf{J}_{-r}=\mathbf{J}^\mathrm{H}_r$. It can be used to shift the target echo at the first range position and obtain the interference echo at ambiguity range position $r$. Assuming that the noise component $\mathbf{n}$ independent of clutter component $\mathbf{c}$, the component of target after the matched filter can be formulated as 
\begin{align}\label{eq5}
\mathbf{y}^{H}(\mathbf{s},v_{d_T})\mathbf{z}&=\alpha\Vert \mathbf{s} \Vert^2_2+\mathbf{y}^{H}(\mathbf{s},v_{d_T})\mathbf{n}\\ \nonumber
&+\sum_{i=1}^{N_t}\gamma_i\mathbf{y}^{H}(\mathbf{s},v_{d_T})\mathbf{J}_{r_i}\mathbf{c}.
\end{align}
Since has been discussed in \cite{20}, shaping STAF is equivalent to minimizing the disturbance power. After discretizing the normalized Doppler frequency interval $[{-}1/2, 1/2]$ into $N_v$ bins and approximating the statistical expectation with sample mean, the disturbance power $P_d$ can be given as \cite{4}
\begin{align}\label{eq6}
{P_d} &= \mathbb{E}\left[ {{{\left| {{{\bf{y}}^H}({\bf{s}},{v_{{d_T}}}){\bf{n}} + \sum\limits_{i = 1}^{{N_t}} {{\gamma _i}{{\bf{y}}^H}({\bf{s}},{v_{{d_T}}}){{\bf{J}}_{{r_i}}}{\bf{c}}} } \right|}^2}} \right]\\ \nonumber
 &= \mathbb{E}\left[ {{{\left| {{{\bf{y}}^H}({\bf{s}},{v_{{d_T}}}){\bf{n}}} \right|}^2}} \right] + \mathbb{E}\left[ {{{\left| {\sum\limits_{i = 1}^{{N_t}} {{\gamma _i}{{\bf{y}}^H}({\bf{s}},{v_{{d_T}}}){{\bf{J}}_{{r_i}}}{\bf{c}}} } \right|}^2}} \right]\\ \nonumber
 &= \sum\limits_{r = 1}^K {\sum\limits_{h = 1}^{{N_v}} {p(r,h){{\left\| {\bf{s}} \right\|}^2}{d_{\bf{s}}}(r,{v_h})} }  + \sigma _n^2{\left\| {\bf{s}} \right\|^2}
\end{align}
where ${v_h} =  - 1/2 + h/{N_v}$ represents the discrete normalized Doppler frequency and $p(r,h)$ is interference map for the range-Doppler bin $(r,v_h)$. Generally, the Doppler frequency of the target is set to 0. Then the STAF of the transmitted code $\mathbf{s}$ is defined as
\begin{align}\label{eq7}
d_s(r,v_h)=\frac{1}{\Vert\mathbf{s} \Vert^2_2}|\mathbf{s}^{H}\mathbf{J}_r\mathbf{y}(\mathbf{s},v_h)|^2.
\end{align}
Under CMC (i.e.$|s(k)|=1,k\in\{0,1,...,K-1\}$), the term $\sigma^{2}_n\Vert\mathbf{s} \Vert^2_2$ in (6) is a constant term. Ignoring the constant term and using a one-to-one mapping for each pair $l \in \{ 0,1,...,K{N_v} - 1\} \to(r,h)\in\{0,1,...,K-1\}\times\{0,1,...,N_v-1\}$, the cost function can be given as  
\begin{align}\label{eq8}
  f(\mathbf{s})&= \sum\limits_{r = 0}^{K - 1} {\sum\limits_{h = 0}^{{N_v} - 1} p } (r,h)|{{\mathbf{s}}^{{H}}}{{\mathbf{J}}}_{r}{\mathbf{y}}({\mathbf{s}},{v_h}){|^2}\\\nonumber
  &=\sum_{i=0}^{K-1}\sum_{i=0}^{N_v-1}|\mathbf{s}^{H}\sqrt{p(r,h)}\mathbf{J}_{i}\text{diag}(\mathbf{p}(v_h))\mathbf{s}|^2\\\nonumber  &=\sum_{i=0}^{KN_v-1}|\mathbf{s}^{H}\mathbf{\Phi}_{i}\mathbf{s}|^2 \\\nonumber
  &=\sum_{i=0}^{KN_v-1}\mathbf{s}^{H}\mathbf{\Phi}_{i}\mathbf{s}\mathbf{s}^{H}\mathbf{\Phi}_{i}^{H}\mathbf{s}
\end{align}
where $\mathbf{\Phi_i}=\sqrt{p(r,h)}\mathbf{s}^{H}\mathbf{J}_i\text{diag}(\mathbf{p}(v_h))$ means the matrix related to interference power distribution. Therefore, the optimization problem of shaping STAF under CMC can be described as the following minimization problem:
\begin{equation}\label{eq9}
{\mathcal{P}}_1\left\{
\begin{array}{rcl}
\mathop{\min} \limits_{\mathbf{s}}&& f(\mathbf{s})\\
s.t.&&|\mathbf{s}(k)|=1, k=0,1,...,K-1
\end{array} \right.
\end{equation}
It has been shown that problem $\mathcal{P}_{1}$ is a quartic and non-convex optimization problem. In \cite{26}, an Iterative Sequential Quartic Optimization (ISQO) algorithm was proposed. It has been shown in \cite{26} where they turn the quartic problem into two alternately quadratic optimization problems and convert them to linear optimization problems. Their work lead to an effective solution but suffer from a slow convergence speed. In \cite{5} and \cite{21}, the authors converted this problem into a Riemannian geometry optimization problem and achieved a relatively good result. However, these methods can only converge to a local optimum.

\section{ADPM FRAMEWORK FOR SOLVING $\mathcal{P}_1$}\label{sec3}
In \cite{23}, X. Yu \textit{et al.} first proposed the ADPM framework for solving problems with highly non-convex constraints. This framework can be executed in parallel in each iteration with lower computational complexity for long sequence designs. Therefore, we use the ADPM framework to increase the penalty factor with iteration number to force the penalty term to approach zero for getting a better feasible solution. Specifically, introducing an auxiliary variable $\mathbf{s}_0$ to $\mathcal{P}_1$, the problem can be recast as
\begin{equation}\label{eq10}
\begin{array}{rcl}
\mathop{min} \limits_{\mathbf{s},\mathbf{s}_0}&& f(\mathbf{s})\\
s.t.&&\mathbf{s}=\mathbf{s}_0\\
&&|\mathbf{s}_0(k)|=1, k=0,1,...,K-1.
\end{array}
\end{equation}
The augmented Lagrangian of $\mathcal{P}_1$ under ADPM framework is expressed as 
\begin{align}\label{eq11}
{\mathcal{L}(\mathbf{s},\mathbf{s}_0},\mathbf{u},\rho)=f(\mathbf{s})+\mathcal{R}\{\mathbf{u}^{H}(\mathbf{s}_0-\mathbf{s})\}+\rho/2\Vert \mathbf{s}-\mathbf{s}_0 \Vert^2_2
\end{align}
where $\mathbf{u}\in\mathbb{C}^{K}$ and $\rho>0$ are the multiplier vector and the penalty parameter respectively. Next we denote $\mathbf{s}^{(r)}$,$\mathbf{s}_0^{(r)}$,$\mathbf{u}^{(r)}$,$\rho^{(r)}$ the update of $\mathbf{s}$, $\mathbf{s}_0$, $\mathbf{u}$, $\rho$ at the $\textit{r}$th iteration. The Algorithm 1 concludes the whole procession of the ADPM framework.

\par In the following, we demonstrate how to update $\mathbf{s}$ and $\mathbf{s}_0$ for Problems (11) .
\begin{algorithm}
\caption{ADPM algorithm for solving the problem (11)}
\label{alg1}
\begin{algorithmic}
\REQUIRE \textit{f}($\mathbf{s}$), $\mathbf{s}^{(0)}$, $\mathbf{s}_0^{(0)}$, ${\bf{u}}^{(0)}$, $\rho^{(0)}$, $\delta_{1,c}$, $\delta_{2,c}$ and \textit{w}, where $0<\delta_{1,c}<1$ and $\delta_{2,c}>1$ are approach to 1, and \textit{w} is a enough large positive number.
\ENSURE An optimized solution $\mathbf{s}^{(*)}$ to Problem (11).
\STATE 1: $r = 1$
\STATE 2: $r = r + 1$
\STATE 3: Upgrade $\mathbf{s}_0^{(r)}$ and $\mathbf{s}^{(r)}$ by solving the following problems,
\begin{equation}\label{eq12}
\begin{array}{l}
\begin{array}{*{20}{c}}
{\begin{array}{*{20}{l}}
{\mathop {\min }\limits_{{{\bf{s}}_0}} }
\end{array}}&{{\mathcal {L}}({{\bf{s}}^{(r - 1)}},{{\bf{s}}_0},{{\bf{u}}^{(r - 1)}},{\rho ^{(r - 1)}})}
\end{array}\\
\begin{array}{*{20}{c}}
{s.t}&{\left| {{{\bf{s}}_0}(k)} \right| = 1,k \in \mathbb{N}},
\end{array}
\end{array}
\end{equation} and
\begin{equation}\label{eq13}
\begin{array}{*{20}{c}}
{\begin{array}{*{20}{l}}
{\mathop {\min }\limits_{\bf{s}} }
\end{array}}&{{\mathcal {L}}({\bf{s}},{\bf{s}}_0^{(r)},{{\bf{u}}^{(r - 1)}},{\rho ^{(r - 1)}})}.
\end{array}
\end{equation}
\STATE 4: Upgrade $\rho^{(r)}$ and $\mathbf{u}^{(r)}$ by
\begin{equation}\label{eq14}
{\rho ^{(r)}} = \left\{ {\begin{array}{*{20}{c}}
{{\rho ^{(r - 1)}}}&{\Delta {r^{(r)}} \le {\delta _{1,c}}\Delta {r^{(r - 1)}}},\\
{{\rho ^{(r - 1)}}{\delta _{2,c}}}&{else,}
\end{array}} \right.
\end{equation}
\begin{equation}\label{eq15}
{\mathbf{u}^{(r)}} = \left\{ {\begin{array}{*{20}{c}}
{\;{\mathbf{\bar{u}}^{(r)}}}&{{u}_{\max }^{(r)} \le w},\\
{{\mathbf{\bar{u}}^{(r)}}/u_{\max }^{(r)}}&{else,}
\end{array}} \right.
\end{equation}
where ${{\bf{\bar{u}}}^{(r)}} = {{\bf{u}}^{(r - 1)}} + {\rho ^{(r)}}({\bf{s}}_0^{(r)} - {{\bf{s}}^{(r)}})$,\\
$u_{\max }^{(r)} = \max \left[ {\left| {\mathbf{\bar{u}}^{(r)}(0)} \right|,...,\left| {{\mathbf{\bar{u}}^{(r)}}(K-1)} \right|} \right]$,\\
$\Delta {r^{(t)}} = \left\| {{\bf{s}}_0^{(r)} - {{\bf{s}}^{(r)}}} \right\|$.
\STATE 5: If the prescribed exit condition is satisfied, output $\mathbf{s}^{(*)}=\mathbf{s}^{(r)}$. Otherwise, return to Step 2.
\end{algorithmic}
\end{algorithm}
\subsection{The optimal upgrade of $\mathbf{s}_0$}\label{sec3.1}
{Upgrade $\mathbf{s}_0$ by solving (12):}
Fix $\mathbf{s}^{(r-1)}$, $\mathbf{s}_0^{(r-1)}$, $\mathbf{u}^{(r-1)}$, $\rho^{(r-1)}$ and upgrade $\mathbf{s}_0$. Ignoring to the irrelevant constant term about $\mathbf{s}_0$ in $\mathcal{L}$($\mathbf{s}$,$\mathbf{s}_0$,$\mathbf{u}$,$\rho$), the problem becomes:
\begin{equation}\label{eq16}
\begin{array}{rcl}
{\mathop {\min }\limits_{{{\bf{s}}_0}} }&&{{\mathcal {R}}\{{{({{\bf{u}}^{\left( {r - 1} \right)}} - {\rho ^{\left( {r - 1} \right)}}{{\bf{s}}^{\left( {r - 1} \right)}})}^{H}}{{\bf{s}}_0}}\}\\
{s.t}&&{\left| {\mathbf{s}_0(k)} \right| = 1,k \in \mathbb{N}}.
\end{array} 
\end{equation}
It is easily observed that problem (16) can be separated into K sub-problems because of the detachable cost function with regard to $\mathbf{s}_0^{(k)}$. When only focusing on 
\textit{k}th problem about $\mathbf{s}_0^{(k)}$ in $\mathbf{s}_0^{(k)}$, it can be expressed as 
\begin{equation}\label{eq17}
\begin{array}{rcl}
{\mathop {\max }\limits_{{{\bf{s}}_0}} }&{{\mathcal {R}}\{ \text\{{\bf{c}}^{(r)}}{{(k)}\}^*{{\bf{s}}_0}(k)\} }\\
{s.t}&{\left| {\mathbf{s}_0(k)} \right| = 1,k \in \mathbb{N}}
\end{array}
\end{equation}
where ${{\bf{c}}^{(r)}} = {\rho ^{(r - 1)}}{{\bf{s}}^{(r - 1)}} - {{\bf{u}}^{(r - 1)}}$.
Problem (17) can be further denoted as:
\begin{equation}\label{eq18}
\begin{array}{*{20}{c}}
{\mathop {\max }\limits_{{\varphi _k}} }&{\cos ({\varphi _k} - {\phi _k})}
\end{array}
\end{equation}
where $\varphi_k$ and $\phi _k$ are the phases of $\mathbf{s}_0^{(k)}$ and $\mathbf{c}_0^{(r)}(k)$.
Therefore, the close-form solution of Problem (18) is:
\begin{equation}\label{eq19}
\begin{array}{rcl}
\varphi _k^* = {\phi _k} 
\end{array}
\end{equation}
\subsection{Upgrading $\mathbf{s}$ with Quartic Riemannian Trust Region Algorithm}
\label{sec3.2}
Fix $\mathbf{s}_0^{(r)}$,$\mathbf{u}^{(r-1)}$,$\rho^{(r-1)}$ and upgrade $\mathbf{s}$. Similarly, ignoring the irrelevant constant term about $\mathbf{s}$ in $\mathcal{L}$($\mathbf{s}$,$\mathbf{s}_0$,$\mathbf{u}$,$\rho$), the problem can be reformed as 
\begin{align}\label{eq20}
{\begin{array}{*{20}{l}}
{\begin{array}{*{20}{c}}
{\min }&{h({\bf{s}}) = f({\bf{s}}) + \frac{1}{2}{\rho ^{(r - 1)}}{{\left\| {{\bf{s}} - {{\bf{v}}^{(r)}}} \right\|}^2}}
\end{array}}\\
{\begin{array}{*{20}{c}}
{s.t.}&{\left| {{\bf{s}}(k)} \right| = 1,k \in \mathbb{N} }
\end{array}}
\end{array}}
\end{align}
where ${{\bf{v}}^{(r)}} = {\bf{s}}_0^{(r)} + {{\bf{u}}^{(r - 1)}}/{\rho ^{(r - 1)}}$.
\begin{proof}
According to problem (12), when fixing $\mathbf{s}_0^{(r)}$,$\mathbf{u}^{(r-1)}$,$\rho^{(r-1)}$ and ignoring the irrelevant items about $\mathbf{s}$, the rest of problem (12) can be recast as 
\begin{align}\label{eq21}
h(\bf{s})&= f({\bf{s}}) - {\mathcal R}\{ ({{\bf{u}}^{(r-1)}})^{{H}}{\bf{s}} 
+ \rho^{(r-1)} ({\bf{s}}_0^{(r)})^{{H}}{\bf{s}} \\\nonumber
&+ \frac{1}{2}\rho^{(r-1)}{\left\| {\bf{s}} \right\|^2}\} \\\nonumber
&= f({\bf{s}}) - {\mathcal R}\{ [({{\bf{u}}^{(r-1)}})^{{H}} 
+ \rho^{(r-1)} {(\bf{s}}_0^{(r)})^{{H}}]{\bf{s}}\}\\\nonumber
&+\frac{1}{2}\rho^{(r-1)}{\left\| {\bf{s}} \right\|^2}\\\nonumber
&= f({\bf{s}}) - {\rho ^{(r - 1)}}{\mathcal R}\{{\bf{v}}^{(r)} {\bf{s}}\}+\frac{1}{2}\rho^{(r-1)}{\left\| {\bf{s}} \right\|^2}
\end{align}
where ${{\bf{v}}^{(r)}} = {\bf{s}}_0^{(r)} + {{\bf{u}}^{(r - 1)}}/{\rho ^{(r - 1)}}$. Considering the collocation method in quadratic polynomial, the equation (21) can be converted to 
\begin{align}\label{eq22}
{h({\bf{s}}) = f({\bf{s}}) + \frac{1}{2}{\rho ^{(r - 1)}}\left\| {{\bf{s}} - {{\bf{v}}^{({r})}}} \right\|^2}-\frac{1}{2} \rho ^{(r - 1)}{\left\| {\bf{v}^{(r)}} \right\|^2}
\end{align}\
The vector ${\bf{v}}$ does not contain a item related to ${\bf{s}}$ and thus the last item of (22) can be omitted. Therefore, the remaining part of (22) can be written as the upper equation of (20).
\end{proof}. 
\par Problem (20) exhibits that it's still a quartic optimization problem with constant modulus restriction. To solve it directly, Riemannian geometry methods such as Quartic Gradient Descent (QGD) algorithm \cite{5} and Riemannian Trust Region (RTR) algorithm \cite{21} was proposed. In QGD algorithm \cite{5}, the author regards CMC as a complex circle manifold and employs the Riemannian gradient descent algorithm to minimize the quartic function on the manifold. Due to only first-order information was used, this algorithm can only achieve linear convergence speed. Since RTR algorithm in \cite{21} can reach a speed of super-linear convergence in non-convex optimization, we here introduce the RTR algorithm to solve the above problem.
\subsubsection{Complex Circle Manifold $\mathcal{M}^{K}$}
\ 
\newline
\indent \\
Considering the CMC as a feasible region on the manifold, problem (20) can be established an unconstrained optimization problem with a K-dimension complex circle Riemannian space 
\begin{align}\label{eq23}
\left\{ {\begin{array}{*{20}{c}}
{\begin{array}{*{20}{c}}
{\mathop {\min }\limits_{\bf{s}} }&{h({\bf{s}})}
\end{array}}\\
{\begin{array}{*{20}{c}}
{s.t.}&{{\bf{s}} \in {{\mathcal M}^K}}
\end{array}}
\end{array}} \right.
\end{align}
where 
\begin{align}\label{eq24}
{\mathcal {M}}^{K} = \{ \mathbf{s} \in {\mathbb{C}}^{K}:\left| {\mathbf{s}(0)} \right| = \left| {\mathbf{s}(1)} \right| = ...\left| {\mathbf{s}(K-1)} \right| = 1\}
\end{align}
 is a product of K-dimension complex manifold as a feasible set (illustrated in Fig.2).i.e.
\begin{figure*}[t]
\centering
\includegraphics[width=6.4in]{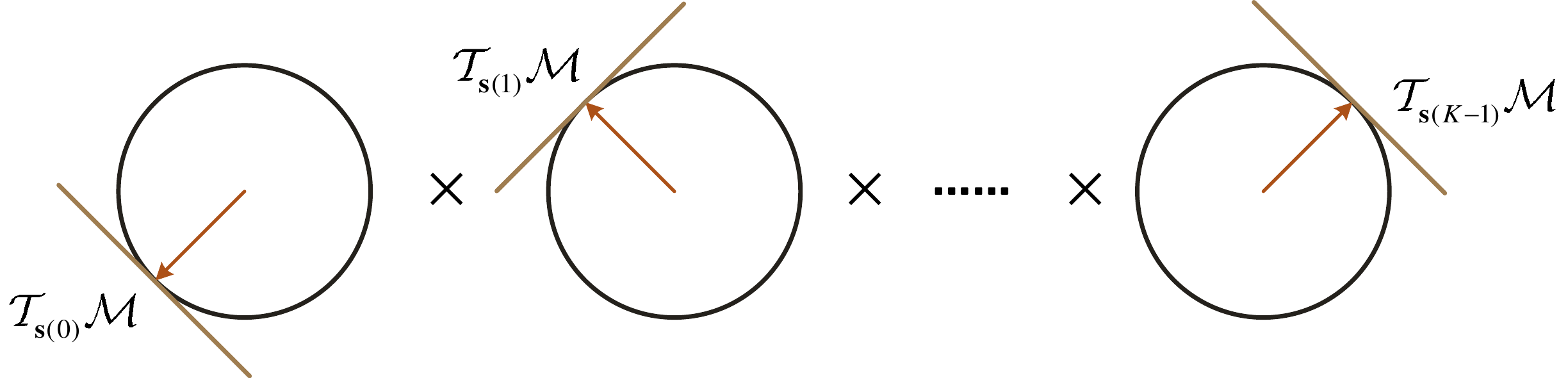}
\caption{K-dimension complex circles product manifold and its tangent space}
\end{figure*}
\begin{align}\label{eq25}{{\mathcal M}^K} = \underbrace {{\mathcal M} \times {\mathcal M} \times ... \times {\mathcal M.}}_{\begin{array}{*{20}{c}}
K&{{\rm{times}}}
\end{array}}
\end{align}
Every ${\mathcal M} \buildrel \Delta \over = \{ x \in \mathbb{C}:{x^*}x = {\mathcal R}{\{ x\} ^2} + {\mathcal I}{\{ x\} ^2} = 1\}$ can be regarded as a sub-manifold of $\mathbb{C}$.

According to \cite{5}, each feasible point $\bf{s}$ (unit modulus point) can be considered as a point in $\mathbb{R}^2$ , by corresponding its real and imaginary parts to the first and second coordinates on the unit circle, i.e.

\begin{align}\label{eq26}
{\bf{s}} &= {\left[ {{\bf{s}}(0),...,{\bf{s}}(k),...,{\bf{s}}(K-1)} \right]^T} \in {{\mathcal M}^K}\\\nonumber &\Rightarrow \left[ {\begin{array}{*{20}{c}}
{{\mathcal R}({\bf{s}}(k))}\\
{{\mathcal I}({\bf{s}}(k))}
\end{array}} \right] \in {\mathcal M}, k = 0,1,...,K - 1.
\end{align}

For each point $x\in\mathcal{M}$, if  manifold $\mathcal {M}$ is a surface on Euclidean space, then the tangent space is the tangent plane of $x$. Therefore, the definition of tangent space is 
\begin{align}\label{eq27}
{{\mathcal T}_x}{\mathcal M} = \{ \bar x \in {{\mathcal T}_{x}}{\mathcal M} = \bar x \in{\mathbb{C}} :{\mathcal R}\{ {x^*}\bar x\}  = 0\}.
\end{align}
Then the tangent space of each ${\bf{s}} \in {{\mathcal M}^K}$, is also the product of K-dimension complex circular tangent spaces
\begin{align}\label{eq28}
{{\mathcal T}_{\bf{s}}}{{\mathcal M}^K} &= {{\mathcal T}_{{\bf{s}}(0)}}{\mathcal M} \times {{\mathcal T}_{{\bf{s}}(1)}}{\mathcal M} \times ... \times {{\mathcal T}_{{\bf{s}}(K - 1)}}{\mathcal M} \\\nonumber
&= \{ {\varsigma _{\bf{s}}} \in {\mathbb{C}}{^K}:{\mathcal R}\{ {{\bf{s}}^*} \odot {\varsigma _{\bf{s}}}\}  = {\bf{0}}\} 
\end{align}
where $\varsigma _{\bf{s}}$ denote the tangent vector in tangent space ${{\mathcal T}_{\bf{s}}}{{\mathcal M}^K}$. For each $\mathbf{s}\in \mathcal {M}$, we use the standardized complex Euclidean inner product to define the Riemannian metric \cite{27}. This metric can be denoted as
\begin{align}\label{eq29}
{g_{\bf{s}}}({\varsigma _{\bf{s}}},{\eta _{\bf{s}}}) = \left\langle {{\varsigma _{\bf{s}}},{\eta _{\bf{s}}}} \right\rangle  = {\mathcal {R}}\left\{ {\varsigma _{\bf{s}}^H,{\eta _{\bf{s}}}} \right\}
\end{align}
where $\varsigma _{\bf{s}}$ and $\eta _{\bf{s}}$ are the tangent vectors in tangent space ${{\mathcal T}_{\bf{s}}}{{\mathcal M}^K}$ with $({\varsigma _{\bf{s}}},{\eta _{\bf{s}}}) \in {{\mathcal T}_{\bf{s}}}{{\mathcal M}^K}$. Thus, the manifold $\mathcal {M}^{K}$ is transferred into a Riemannian sub-manifold of the Euclidean space $\mathbb{C}^K $ with this kind of metric. Furthermore, in order to keep the iteration points on the manifold $\mathcal {M}^{K}$ throughout the gradient descent, a retraction mapping from ${{\mathcal T}_{\bf{s}}}{{\mathcal M}^K}$ to $\mathcal {M}^{K}$ is defined as \cite{27}.
\begin{align}\label{eq30}
{R_{\bf{s}}}({\varsigma _{\bf{s}}}) \buildrel \Delta \over = \frac{{{\bf{s}} + {\varsigma _{\bf{s}}}}}{{\left\| {{\bf{s}} + {\varsigma _{\bf{s}}}} \right\|}}.
\end{align}
For complex circle manifold, the orthogonal projector is given by 
\begin{align}\label{eq31}
{\rm{Pro}}{{\rm{j}}_{\bf{s}}}({\varsigma _{\bf{s}}}) = {\varsigma _{\bf{s}}} - \left\langle {{\bf{s}},{\varsigma _{\bf{s}}}} \right\rangle {\bf{s}}.
\end{align}
According to \cite{27}, the gradient of Riemannian in ${{\mathcal T}_{\bf{s}}}{{\mathcal M}^K}$ is defined as 
\begin{align}\label{eq32}
{g_{\bf{s}}}({\varsigma _{\bf{s}}},{\rm{grad}}h(s)) = Df(\bf{s})\left[ {{\varsigma _{\bf{s}}}} \right].
\end{align}
where $Df({\bf{s}})\left[ {{\varsigma _{\bf{s}}}} \right] \buildrel \Delta \over = \frac{1}{a} \mathop {\lim }\limits_{a \to 0} h({\bf{s}} + a{\varsigma _{\bf{s}}}) - h({\bf{s}})$ denotes the directional derivative of the cost function $h({\bf{s}})$ at point $\bf{s}$ along direction $\varsigma _{\bf{s}} \in {{\mathcal T}_{\bf{s}}}{{\mathcal M}^K}$. Specifically, ${\rm{grad}}h({\bf{s}})$ can be calculated by projecting the Euclidean gradient of $h({\bf{s}})$ onto the tangent space ${{\mathcal T}_{\bf{s}}}{{\mathcal M}^K}$ \cite{28}
\begin{align}\label{eq33}
{\rm{grad}}h({\bf{s}})&={\rm{Pro}}{{\rm{j}}_{\bf{s}}}({\rm{Grad}}h({\bf{s}})) \\\nonumber &={\rm{Grad}}h({\bf{s}}) - {\mathcal {R}}\{ {\rm{Grad}}h({\bf{s}}) \odot {\bf{s}^*}\}  \odot {\bf{s}},
\end{align}
where ${\rm{Grad}}h({\bf{s}})$ Euclidean gradient of $h({\bf{s}})$. Similarly, Riemannian Hessian matrix ${\rm{hess}} h({\bf{s}})\left[ {{\varsigma _{\bf{s}}}} \right]$ of $h({\bf{s}})$ on the tangent space, which is calculated by \cite{28}
\begin{align}\label{eq34}
{\rm{hess}}h({\bf{s}})\left[{{\varsigma _{\bf{s}}}} \right]&={\rm{Pro}}{{\rm{j}}_{\bf{s}}}(D{\rm{grad}}h({\bf{s}})\left[{{\varsigma _{\bf{s}}}} \right])\\\nonumber
 &= {\rm{Pro}}{{\rm{j}}_{\bf{s}}}({\rm{Hess}}h({\bf{s}})\left[ {{\varsigma _{\bf{s}}}} \right]) - \left\langle {{\bf{s}},{\rm{grad}}h({\bf{s}})} \right\rangle {\varsigma _{\bf{s}}}
\end{align}
where ${\rm{Hess}}h({\bf{s}})$ is the Euclidean Hessian matrix.

The specific introduction of derivation of RTR algorithm as follow.
\subsubsection{RTR algorithm and derivation for complex circle manifold}
\ 
\newline
\indent \\
Giving a function $f$, iteration variables $x \in \mathbb{R}^{n}$ and update variables $\theta \in \mathbb{R}^{n}$ on a Euclidean space, the trust region method for the Euclidean space $\mathbb{R}$ function $f$ requires solving the following trust region sub-problem \cite{29}
\begin{align}\label{eq35}
\begin{array}{*{20}{lll}}
\begin{array}{*{20}{c}}
{\mathop {\min }\limits_{\theta  \in \mathbb{R}^{n}} {m_x}(\theta ) = f(x) + {\theta ^T}\partial f(x) + \frac{1}{2}{\theta ^T}{\partial ^2}f(x)\theta ,}&{\left\| \theta  \right\| \le \Delta }
\end{array}
\end{array}
\end{align}
where $\partial {f(x)}$ is the Euclidean space gradient of the function $f$ at $x \in \mathbb{R}^n$. Similarly, ${\partial ^2}f(x)$ is the Hessian matrix of the Euclidean space of $f$ at $x \in \mathbb{R}^n$, and $\Delta $ denotes the radius of trust region.

Based on the contraction property in Riemannian geometry, given the objective function $ h({\bf{s}}):{{\mathcal M}^K} \to \mathbb{C}$ and the iteration variable a. Combining the retraction operator in (30) and the Riemannian metric in (32), define the trust region sub-problem for the tangent space ${{\mathcal T}_{\bf{s}}}{{\mathcal M}^K}$
\begin{flalign}\label{eq36}
\mathop {\min }\limits_{{{\bf{\varsigma }}_k} \in {{\mathcal T}_{\bf{s}}}{{\mathcal M}^K}} {m_k}({{\bf{\varsigma }}_k}) &= {{\hat h}_k}({\bf{0}}) + D{{\hat h}_k}({\bf{0}})[{{\bf{\varsigma }}_k}]\\\nonumber
&+ \frac{1}{2}{D^2}{{\hat h}_k}({\bf{0}})[{{\bf{\varsigma }}_k},{{\bf{\varsigma }}_k}]\\\nonumber
&= {{\hat h}_k}({\bf{0}}) + \left\langle {{\rm{grad }}{{\hat h}_k}({\bf{0}}),{{\bf{\varsigma }}_k}} \right\rangle\\\nonumber
&+ \frac{1}{2}\left\langle {{\rm{hess }}{{\hat h}_k}({\bf{0}})\left[ {{{\bf{\varsigma }}_{\bf{s}}}} \right],{{\bf{\varsigma }}_k}} \right\rangle \\\nonumber
\begin{array}{*{20}{l}}
{s.t.}&{\left\langle {{{\bf{\varsigma }}_{\bf{s}}},{{\bf{\varsigma }}_{\bf{s}}}} \right\rangle }
\end{array} \le \Delta _k^2
\end{flalign}
where ${\hat h_k}({\bf{0}}):{{\mathcal T}_{\bf{s}}}{{\mathcal M}^K} \to C:{{\bf{\varsigma }}_k} \to f(R({{\bf{\varsigma }}_k}))$. According to the definition of retraction, we have
\begin{align}\label{eq37}
\left\{ {\begin{array}{*{20}{c}}
{{{\hat h}_k}({\bf{0}}) = h({{\bf{s}}_k})}\\
{{\rm{grad }}{{\hat h}_k}({\bf{0}}) = {\rm{grad }}h({{\bf{s}}_k})}\\
{{\rm{hess }}{{\hat h}_k}({\bf{0}}) = {\rm{hess }}h({{\bf{s}}_k})}
\end{array}} \right.
\end{align}
Considering the quadratic term symmetric form, equation (36) can be written as
\begin{flalign}\label{eq38}
\mathop {\min }\limits_{{{\bf{\varsigma }}_k} \in {{\mathcal T}_{\bf{s}}}{{\mathcal M}^K}} {m_k}({{\bf{\varsigma }}_k}) &= h({{\bf{s}}_k}) + \left\langle {{\rm{grad }}h({{\bf{s}}_k}),{{\bf{\varsigma }}_k}} \right\rangle \\\nonumber
 &+ \frac{1}{2}\left\langle {{\rm{hess }}h({{\bf{s}}_k})\left[ {{{\bf{\varsigma }}_{\bf{s}}}} \right],{{\bf{\varsigma }}_k}} \right\rangle \\\nonumber
\begin{array}{*{20}{l}}
{s.t.}&{\left\langle {{{\bf{\varsigma }}_{\bf{s}}},{{\bf{\varsigma }}_{\bf{s}}}} \right\rangle }
\end{array} \le \Delta _k^2
\end{flalign}
The specific solution procedure for the Riemannian gradient and Riemannian Hessian matrix is shown in Appendix A.

Similar to trust region algorithm in Euclidean space, the points obtained by solving the trust region sub-problem are used as candidates for the next iteration, and these candidates can be accepted or rejected by comparing the quadratic model with the declining similarity degree of the original function. Defining quotient as
\begin{align}\label{eq39}
\begin{array}{lcl}
{\chi _k} = \frac{{h({{\bf{s}}(k)}) - h[{R_{{{\bf{s}}(k)}}}({\varsigma _{{{\bf{s}}(k)}}})]}}{{{m_k}({{\bf{0}}_K}) - {m_k}({\varsigma _{{{\bf{s}}(k)}}})}}.
\end{array}
\end{align}
If $\chi _k$ is very low, the model must be undesirable. This iteration result should be refused  and enlarge the trust region radius $\Delta$. In contrast, when $\chi _k$ is nearly approaching to 1, the ideal iteration result can be accepted and enlarge the trust region radius.

The specific quartic RTR algorithm is summarized in Algorithm 2.
\begin{algorithm}
\caption{Quartic RTR algorithm}
\label{alg2}
\begin{algorithmic}
\REQUIRE The interference map $p(r,h)$, stating point ${{\bf{s}}(0)} \in {{\mathcal {M}}_s}$. {$i=0$}, $\mathop{\bar \Delta}> 0$, ${\Delta _0} \in {(0,\mathop {\bar{\Delta}})} $ and ${\chi ^{'}} \in \left[ {0,\frac{1}{4}} \right)$
\ENSURE An optimized solution ${{\bf{s}}^*} \in {\mathcal {M}}$.
\FOR {$r=1:K-1$}
\FOR {$h=1:N_v-1$}
\STATE Calculate $v_h$, $\mathbf{p}$($v_h$) and $\mathbf{\Phi}_i$;
\STATE {$i=i+1$};
\ENDFOR
\ENDFOR
\FOR {$j=0, 1, ...$}
\STATE Calculate ${\varsigma _{{{\bf{s}}(j)}}}$ by solving(30);
\STATE Calculate ${\chi _j}$ by solving(31);
\IF{$\chi _j<\frac{1}{4}$}
\STATE${\Delta _{j + 1}} = \frac{1}{4}{\Delta _j}$;
\ELSE \IF{$\chi _j>\frac{3}{4}$ and $\left\| {\varsigma _{{\bf{s}}{(j)}}} \right\| = {\Delta _j}$}
\STATE ${\Delta _{j + 1}} = \min (2{\Delta _j},\mathop {\bar{\Delta}})$;
\ELSE
\STATE${\Delta _{j + 1}} = {\Delta _j}$;
\ENDIF
\ENDIF
\IF{$\chi _{k} > \chi ^{'}$}
\STATE ${{\bf{s}}(j + 1)} = {R_{{{\bf{s}}(j)}}}({\varsigma _{{{\bf{s}}(j)}}})$;
\ELSE 
\STATE ${{\bf{s}}(j + 1)} = {{\bf{s}}(j)}$;
\ENDIF
\ENDFOR
\end{algorithmic}
\end{algorithm}
Since the sub-problem is given in Euclidean space, the definition of a Cauchy point can be migrated from Euclidean space and has a lower bound
\begin{align}\label{eq40}
\begin{array}{lcl}
\begin{array}{lcl}
{m_k}({\varsigma _{{{\bf{s}}(k)}}}) \le {m_k}({{\bf{0}}_K})\\
- \kappa \left\| {{\rm{grad}}h({{\bf{s}}(k)})} \right\|\min \left\{ {{\Delta _k},\frac{{\left\| {{\rm{grad}}h({{\bf{s}}(k)})} \right\|}}{{\left\| {{\rm{hess}}({{\bf{s}}(k)})\left[ {{\varsigma _{{{\bf{s}}(k)}}}} \right]} \right\|}}} \right\}
\end{array}
\end{array}
\end{align}
where $\kappa$ is a constant and $\kappa  > 0$. The global convergence can be obtained by applying the concept of Lipschitz continuous differentiability on the complex circular manifold $\mathcal {M}$. According to \cite{28}, the Lipschitz continuous differentiability of the cost function f($\mathbf{s}$) can be proved under the assumptions of $\mathop {\bar{\chi}} \in \left[ {0,\frac{1}{4}} \right)$.

\subsection{Termination Condition}\label{sec3.3}
According to \cite{30}, the ADPM algorithm will converge to a stationary point in the case of continuous phase. The proper termination condition of ADPM algorithm can be expressed as
\begin{align}\label{eq41}
\begin{array}{lll}
{\left\| {{\bf{s}}_0^{(r)} - {{\bf{s}}^{(r)}}} \right\|_2} \le \varepsilon _{pri}^{(r)}
\end{array}
\end{align}
\begin{align}\label{eq42}
\begin{array}{lll}
{\rho ^{(0)}}{\left\| {{\bf{s}}_0^{(r - 1)} - {\bf{s}}_0^{(r)}} \right\|_2} \le \varepsilon _{dual}^{(r)}
\end{array}
\end{align}
where $\varepsilon _{pri}^{(r)}$ and $\varepsilon _{dual}^{(r)}$ represent the feasibility tolerances for primal and dual residuals at the {\textit{r}}th iteration, respectively. These can be calculated by absolute and a relative criteria, such as
\begin{align}\label{eq43}
\begin{array}{lll}
\varepsilon _{pri}^{(r)} = \sqrt {2N} {\varepsilon _{abs}} + {\varepsilon _{rel}}\max \left\{ {\left\| {{{\bf{s}}^{(r)}}} \right\|,\left\| {{\bf{s}}_0^{(r)}} \right\|} \right\}
\end{array}
\end{align}
\begin{align}\label{eq44}
\begin{array}{lll}
\varepsilon _{dual}^{(r)} = \sqrt {2N} {\varepsilon _{abs}} + {\varepsilon _{rel}}\left\| {{\bf{s}}_0^{(r)}} \right\|
\end{array}
\end{align}
where ${\varepsilon _{abs}}$ is the absolute tolerance and ${\varepsilon _{rel}}$ is the relative tolerance.

Moreover, a proper choice of $\rho ^{(0)}$ can reach convergence faster. For this purpose, we use the method proposed in \cite{23} to initialize the penalty factor for ADPM framework. $f(\mathbf{s})$ in problem (11) has no non-convex constraint, then the initialized penalty factor of ADPM can be given by \cite{31}
\begin{align}\label{eq45}
{\rho ^{(0)}} = \sqrt {{\lambda _{\max }}[f(\mathbf{s})]{\lambda _{\min }}[f(\mathbf{s})]} 
\end{align}\
where ${\lambda _{\max }}$ and ${\lambda _{\min }}$ denote the maximum and minimum eigenvalues, respectively. Note that if the smallest eigenvalue is zero, ${\lambda _{\min }}$ is determined by its smallest non-zero eigenvalue. Here, we choose the smallest eigenvalue larger than $10^{-3}$.
\section{Numerical Experiment}
In this section, we provide several numerical experiments to evaluate the performance of the proposed algorithm. We compare the performance with the following approaches: 1) Inexact ADPM with ISQO algorithm; 2) Only Riemannian Trust Region Algorithm. 
\par Considering a radar system with $N_v = 50$, $K = 50$, we set the multiplier vector ${\bf{u}}^{(0)}$ = ${{\mathbf{0}}}_{K\times 1}$, ${\varepsilon _{abs}} = {10^{ - 6}}$, ${\varepsilon _{dual}} = {10^{ - 6}}$,  ${\delta _{1,c}} = 0.97$, ${\delta _{2,c}} = 1.03$ and $w=10^4 $ for whole ADPM framework. In the case of ISQO algorithm in inexact ADPM framework, we set the maximum iteration of ISQO is 30 with the exit condition $\begin{array}{*{20}{l}}
{\left| {f({\mathbf{s}^{(r)}}) - f({\mathbf{s}^{(r - 1)}})} \right|}
\end{array} \le {10^{ - 6}}$.
Considering RTR algorithm, we let the upper bounds on the radius of the trust region $\mathop {\bar \Delta}_0$ and the radius of the initial trust region to  ${(K)^{\frac{1}{2}}}$ and ${\bar \Delta}$/8 respectively, and the exit condition gradient norm less than $10^{-6}$ without any iteration times limit. In our proposal algorithm, the maximum iteration of RTR algorithm in ADPM is setting as 30 with the same termination condition in above all. To prevent the situation that 
$f({\mathbf{s}})$ in equation (20) may appear non-positive, the value of $\rho^{(r)}$ in  \textit{r}th iteration will be halved if $f({\bf{s}}) \le 0$ in one step of the iteration.
We evaluate all numerical assessment by providing signal-to-interference-ratio (SIR) (actually minimizing the cost function values). The SIR is defined as
\begin{align}\label{eq46}
SIR = \frac{{{K^2}}}{{\sum\limits_{r = 1}^K {\sum\limits_{h = 1}^{{N_v}} {p(r,h){{\left\| {\bf{s}} \right\|}^2}{d_{\bf{s}}}(r,{v_h})} } }}.
\end{align}
We use the $P4.^3$ code as the original waveform and the the specific expression is 
\begin{align}\label{eq47}
\begin{array}{lll}
\mathbf{s}(k) = \exp \{ j\pi [(k - 1)^2/K - (k - 1)]\} ,k \in {\mathbb{N}}.
\end{array}
\end{align}
and the STAF of $P4.^3$ code (original STAF) is drawn in Fig. 3
\begin{figure}[htpb]
\centering
\centering
\includegraphics[width=3.2in]{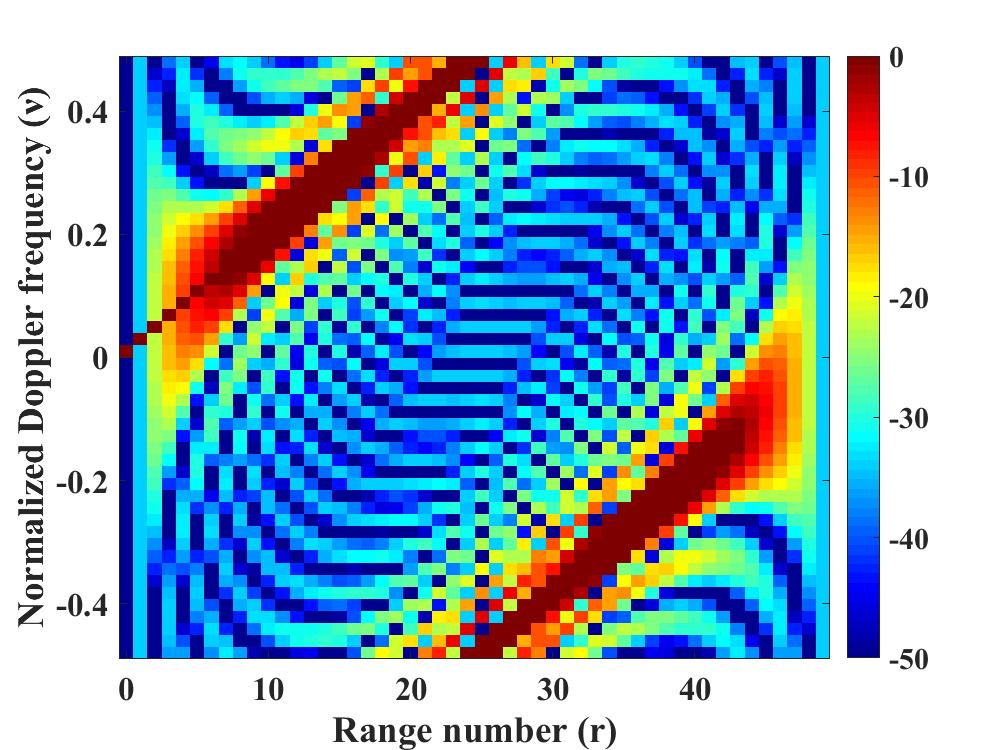}
\caption{Original STAF for $P4.^3$ code.}
\end{figure}

\begin{remark}\label{rem1}
We used the same platform for all experiments: MATLAB 2016b, CPU Core i7-8750H, 2.21 GHz and 8 GB of RAM.
\end{remark}
\subsection{Scene 1: One Interference Area}
\par We consider the following desired STAF. For scene 1, its interference distribution map is setting as
\begin{align}\label{eq48}
\begin{array}{lll}
{p(r,h) = \left\{ {\begin{array}{*{20}{c}}
{\begin{array}{*{20}{c}}
{1,}&{(r,h) \in \{ 18,19,20\}  \times \{ 35,...,47\} }.
\end{array}}\\
{\begin{array}{*{20}{c}}
{0,}&{otherwise}
\end{array}}
\end{array}} \right.}
\end{array}
\end{align}
The STAF of scene 1 can be shown in Fig. 4 
\begin{figure}[htpb]
\centering
\centering
\includegraphics[width=3.2in]{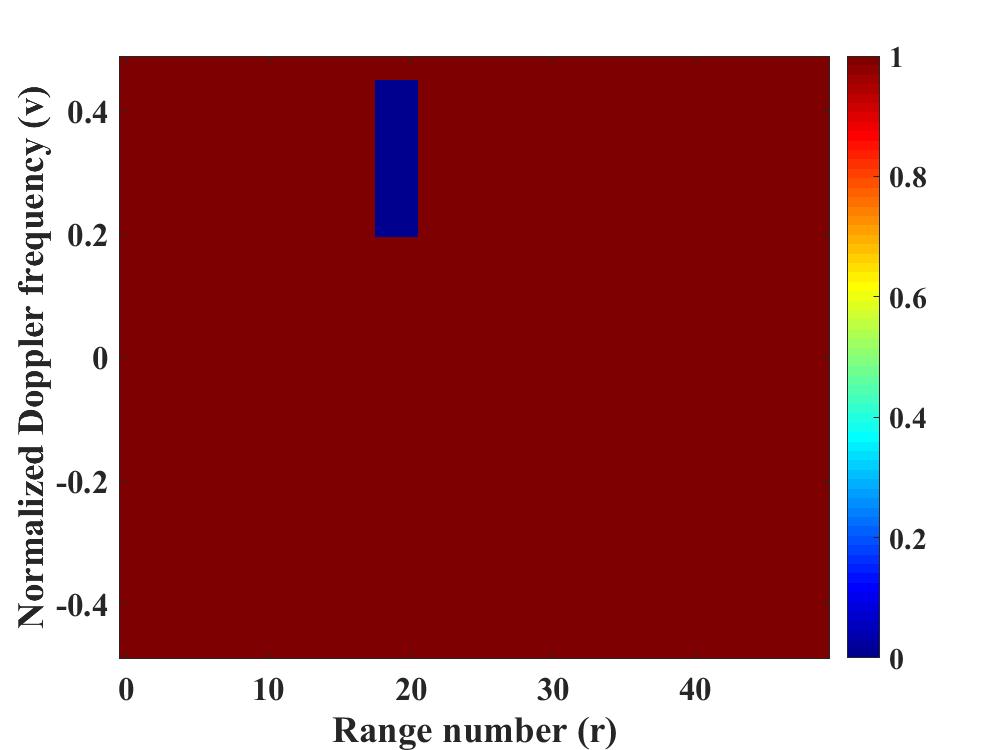}
\caption{Desired STAF for Scene 1.}
\end{figure}

\begin{figure}[htpb]
\centering
\subfigure[]
{
\centering
\includegraphics[width=3.2in]{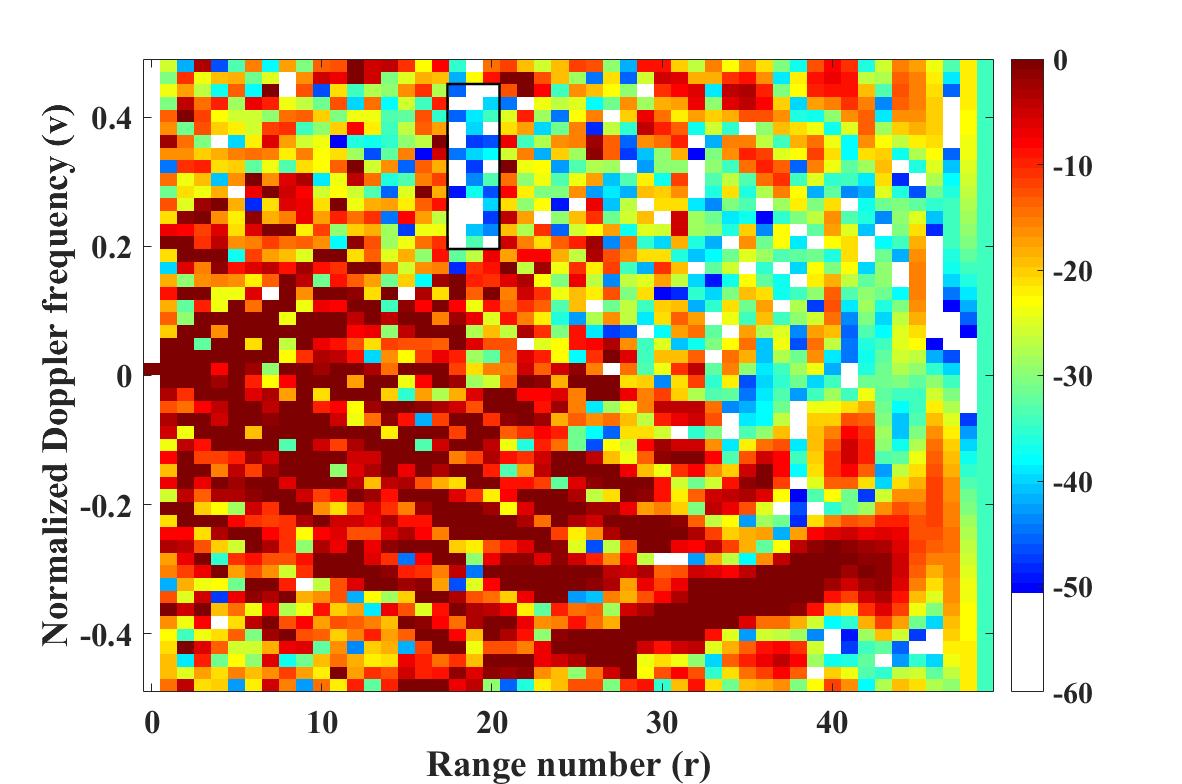}
\quad
}\\
\subfigure[]{
\centering
\includegraphics[width=3.2in]{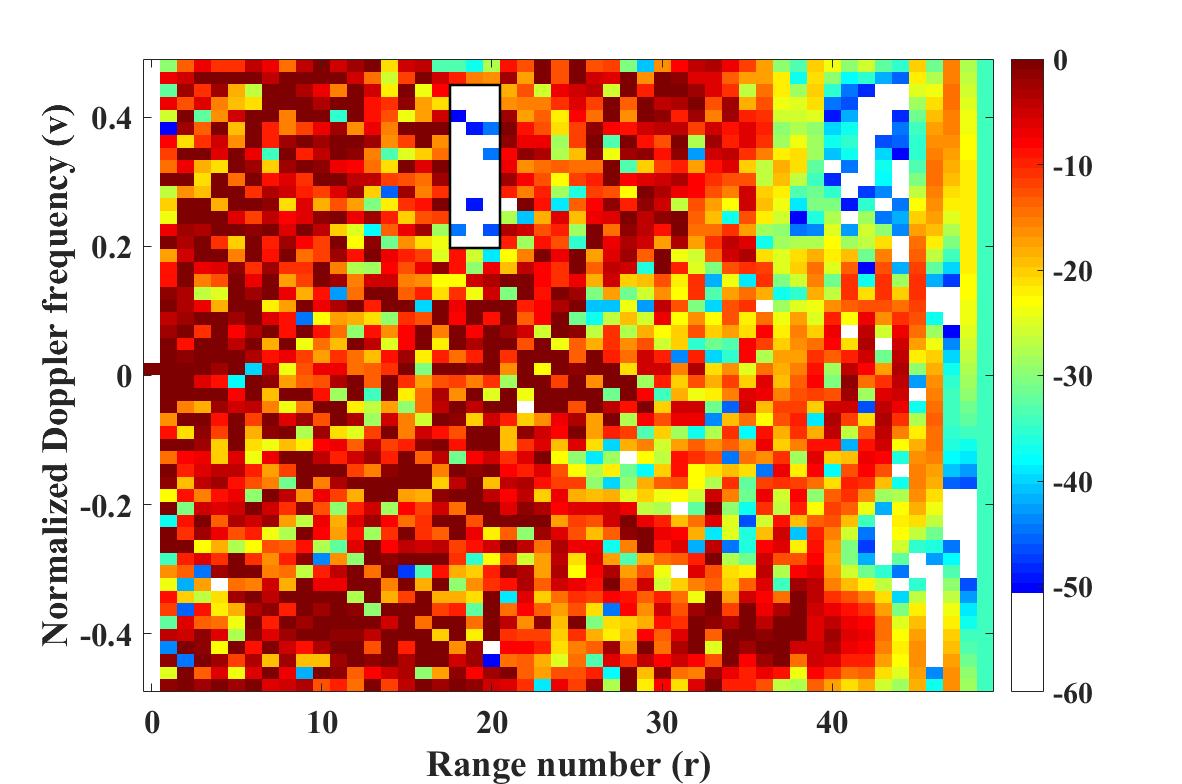}
\quad
}\\
\subfigure[]{
\centering
\includegraphics[width=3.2in]{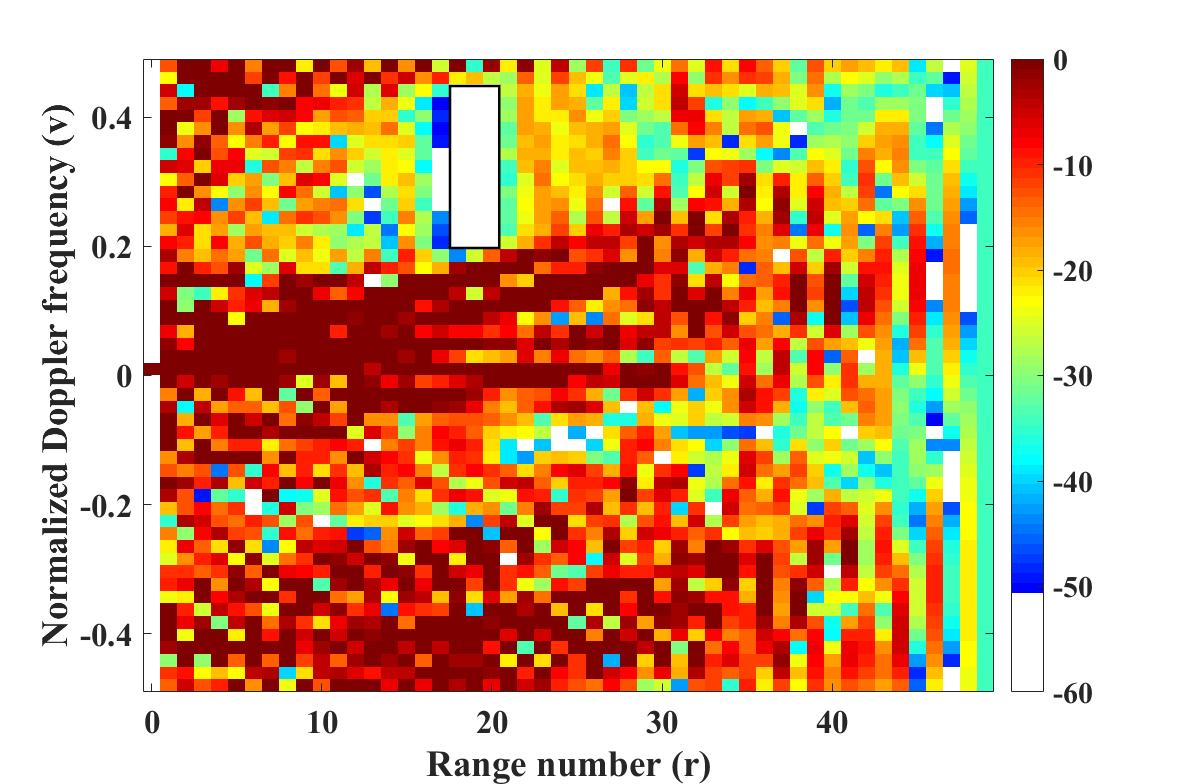}
\quad
}%
\caption{STAFs produced by different algorithms in scene 1: (a) STAF for inexact ADPM+ISQO, (b) STAF for RTR, and (c) STAF for ADPM+RTR.}
\label{3}
\end{figure}
Fig.5 (a)-(c) represent 2-D STAFs based on $P4.^3$ for Inexact ADPM+ISQO, RTR and ADPM+RTR. From these figures, it is worth mentioning that all three methods are shown the ability of suppressing signals in the disturbance area, but it is obvious that the response for the ADPM+RTR waveform is the nearest one to the desired (the black rectangles indicate the areas where the interference needs to be suppressed with average value about -75dB).
\par To further demonstrate the interference suppression effect of the designed radar waveform, we compare the response values for different range-cuts along STAFs. The range-cuts at three different range locations $r = 18, 19, 20$ are provided in Fig. 6(a)–(c), respectively. From the data, it can be seen that all three methods produce reasonable responses along the normalized Doppler frequency axis over the specified region. In most cases, it can be verified that our proposed algorithm comes closest to the desired STAF response.
\begin{figure}[htpb]
\centering
\subfigure[]
{
\centering
\includegraphics[width=1.6in]{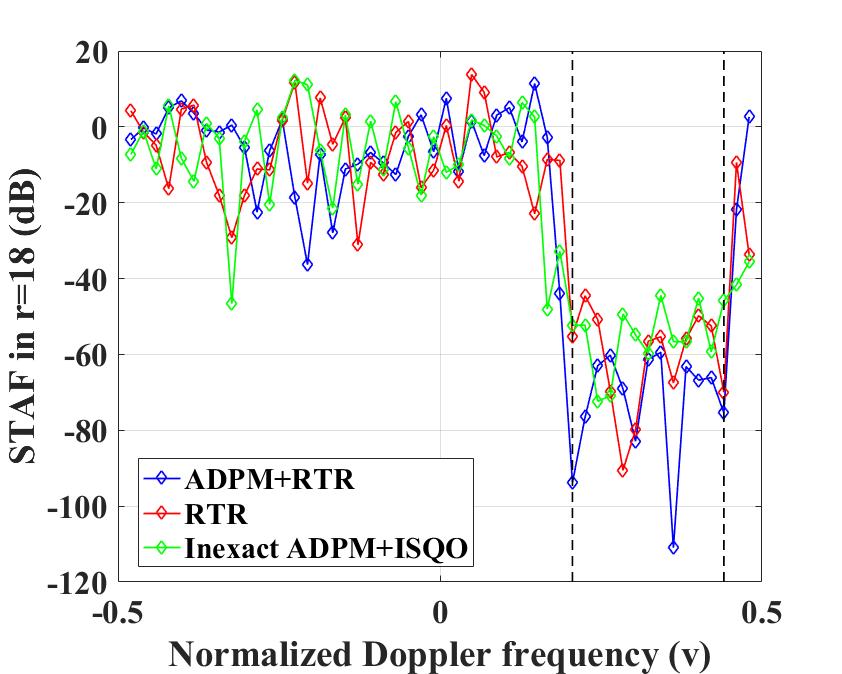}
\quad
}
\subfigure[]{
\centering
\includegraphics[width=1.6in]{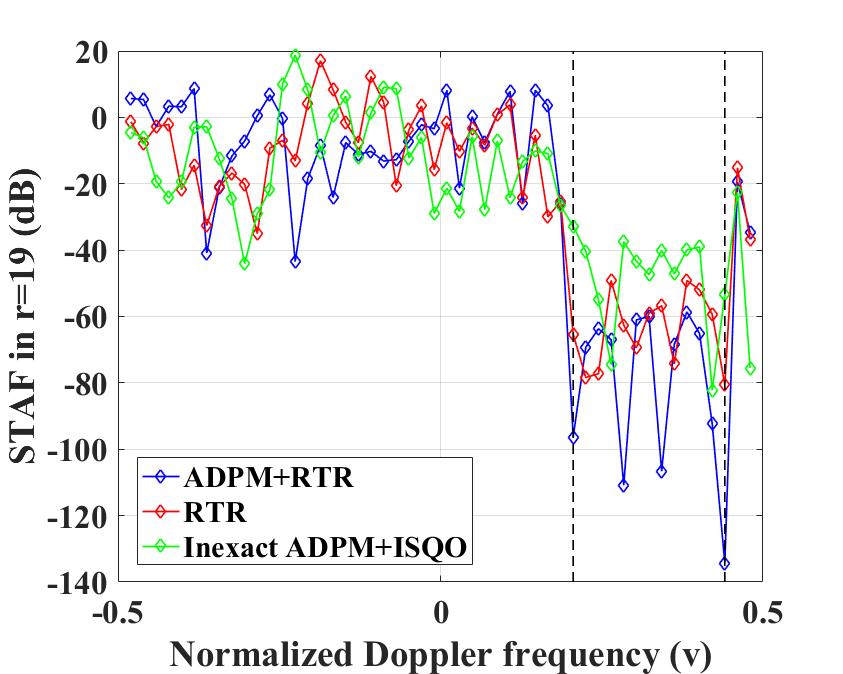}
\quad
}\\
\subfigure[]{
\centering
\includegraphics[width=1.6in]{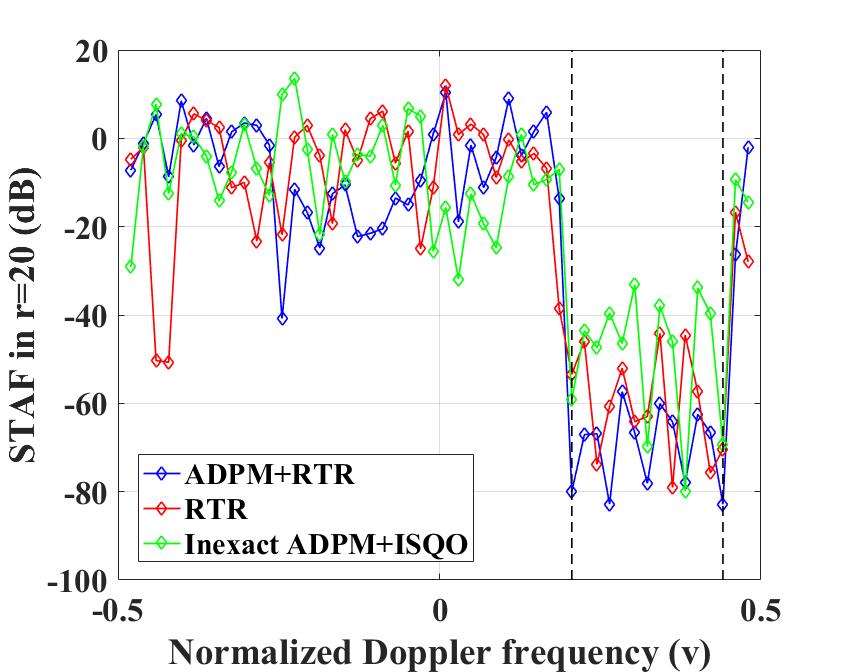}
\quad
}%
\caption{STAF cuts at (a) r = 18, (b) 19, and (c) 20.}
\label{3}
\end{figure}
The achieved SIR average values and the average simulation times for the all competing methods are plotted in Fig. 7. It is clear from these figures that the ADPM+RTR algorithm achieves the highest SIR values. In addition, in terms of the run-time of each method, the ADPM+RTR algorithm takes more time because it requires two-layer iterations.
\begin{figure}[htpb]
\centering
\centering
\includegraphics[width=3.2in]{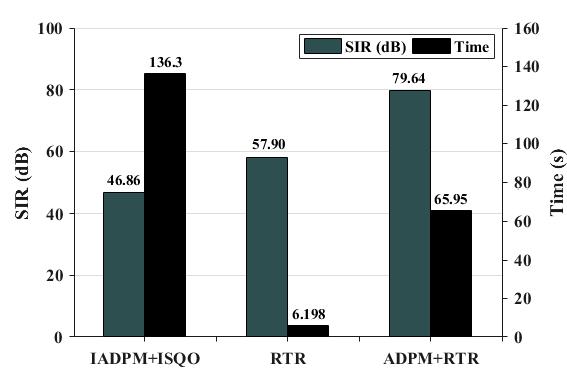}
\caption{SIR average values and average simulation times for Inexact ADPM+ISQO, RTR and ADPM+RTR.}
\end{figure}

\subsection{Scene 2: Two Interference Areas}
The interference distribution map $p(r, h)$ for this scene is defined as (See Fig. 8)
\begin{align}\label{eq49}
p(r,h) = \left\{ {\begin{array}{*{20}{c}}
{1,}\\
{1,}\\
{0,}
\end{array}\begin{array}{*{20}{c}}
{(r,h) \in \{ 20,21,...,30\}  \times \{ 14,15\} }\\
{(r,h) \in \{ 23,24,25\}  \times \{ 38,39,...,42\} }\\
{otherwise}
\end{array}} \right.
\end{align}
\begin{figure}[htpb]
\centering
\centering
\includegraphics[width=3.2in]{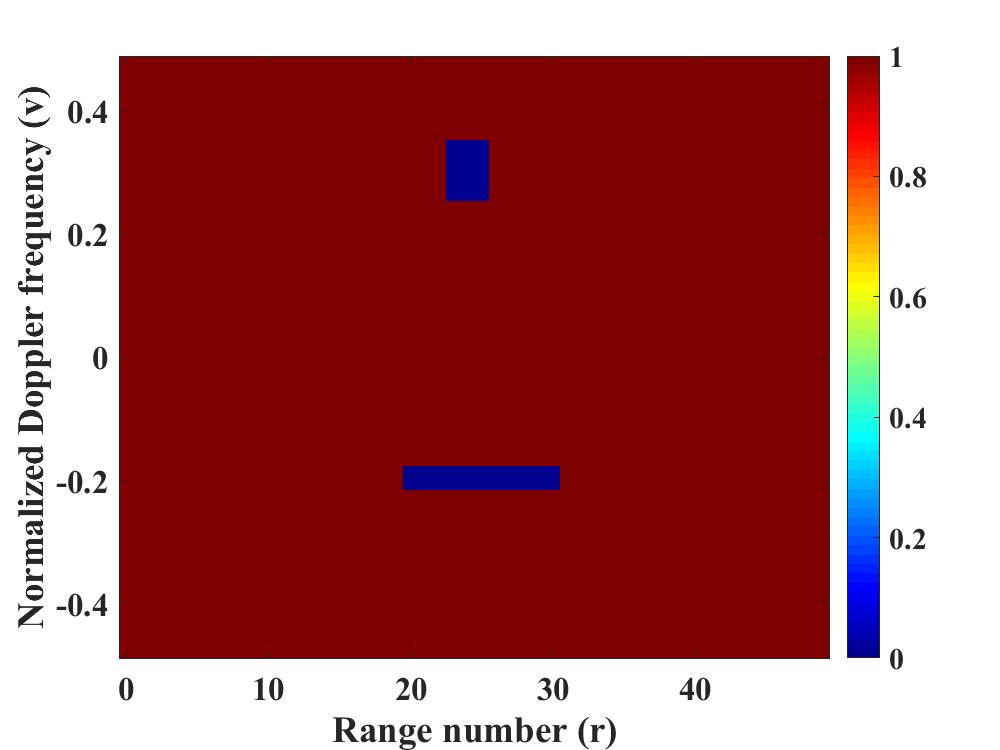}
\caption{Desired STAF for Scene 2.}
\end{figure}

In scene 2, the 2-D figures of the STAF for all methods are provided in Fig. 9(a)-(c) along with the desired response. From these figures, all three algorithms also show the effect of suppressing in designated area and ADPM+RTR algorithm is still the closest one (The black rectangle indicates the area where the interference is suppressed with average value around -67dB). 
\begin{figure}[htpb]
\centering
\subfigure[]
{
\centering
\includegraphics[width=3.2in]{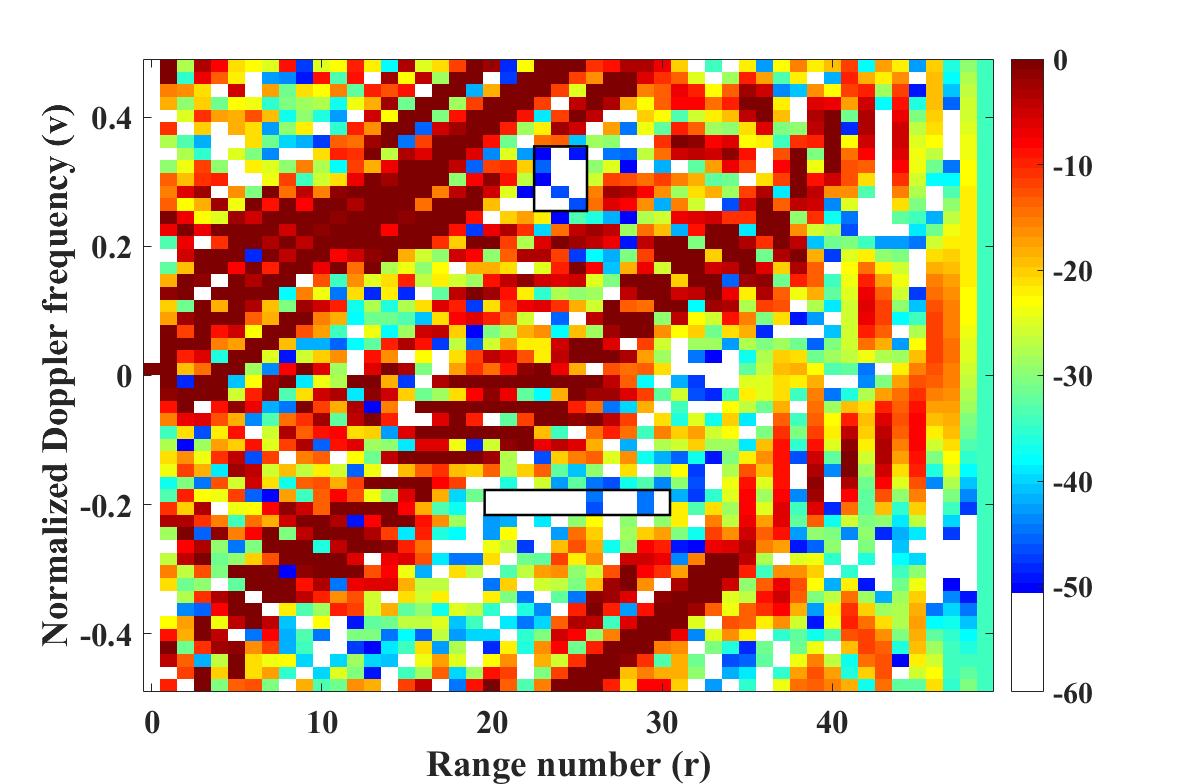}
\quad
}\\
\subfigure[]{
\centering
\includegraphics[width=3.2in]{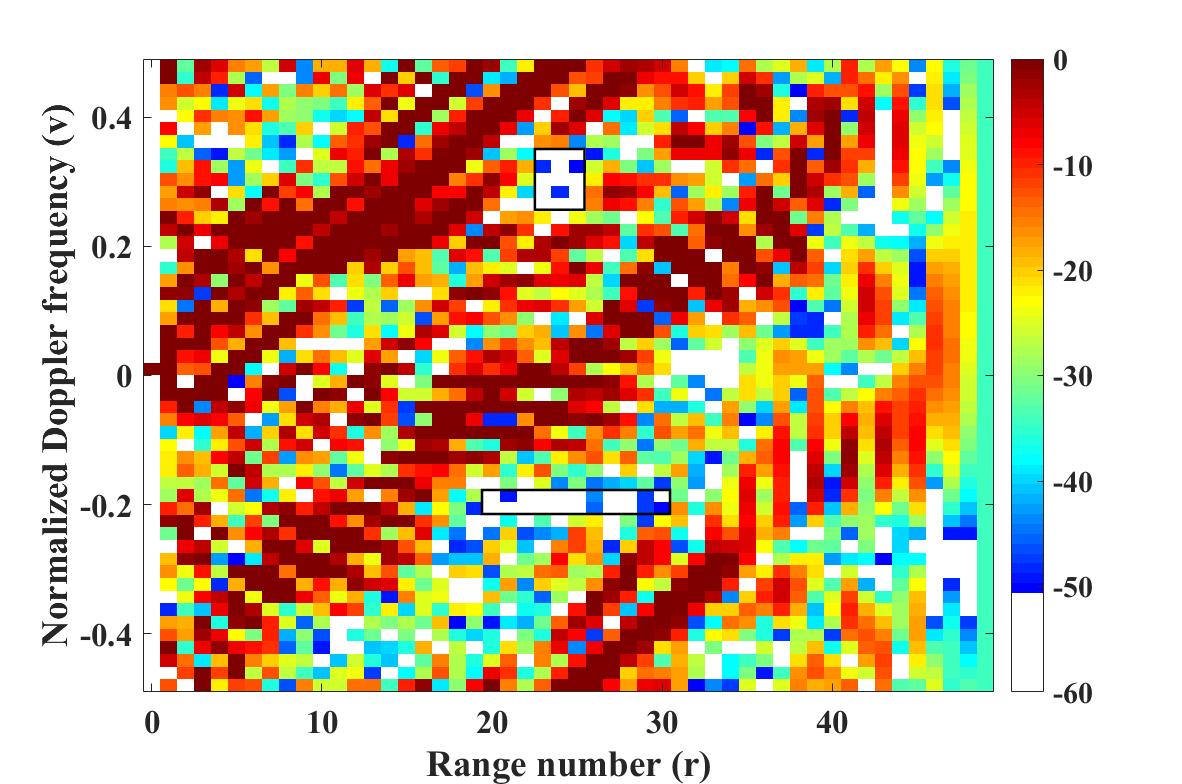}
\quad
}\\
\subfigure[]{
\centering
\includegraphics[width=3.2in]{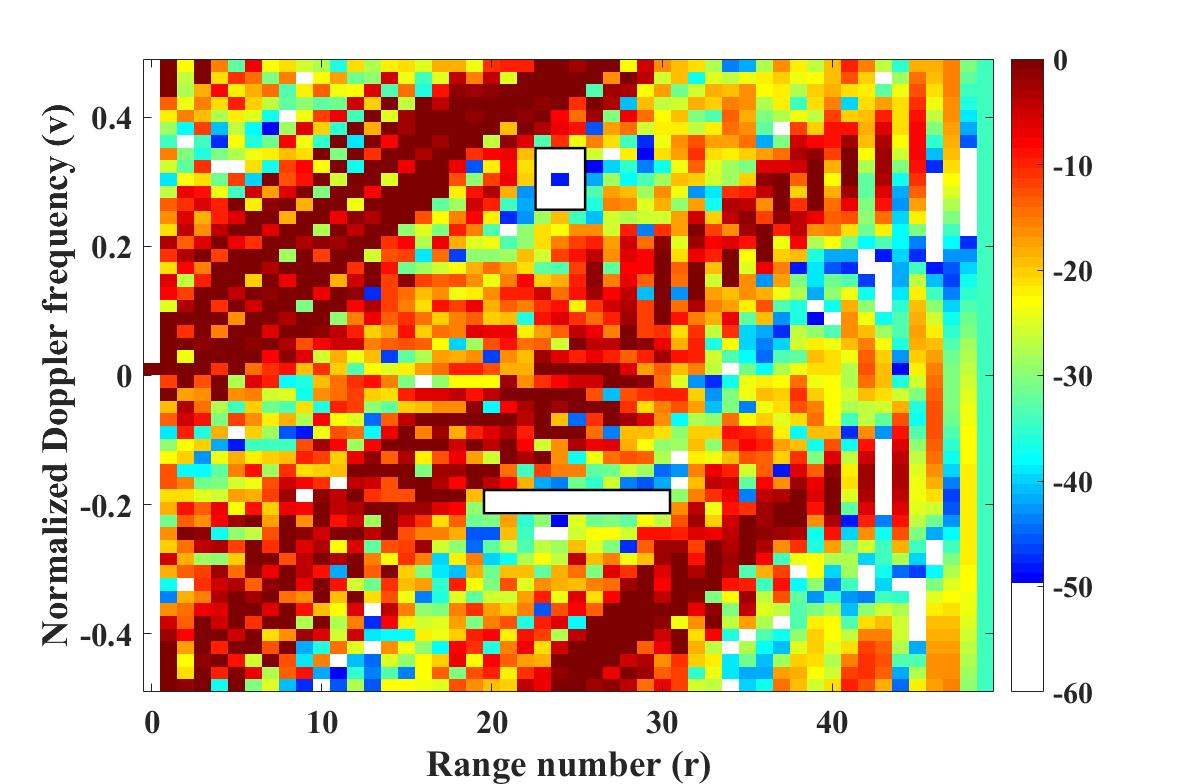}
\quad
}%
\caption{STAFs produced by different algorithms in scene 2: (a) STAF for inexact ADPM+ISQO, (b) STAF for RTR, and (c) STAF for ADPM+RTR.}
\label{3}
\end{figure}
In Fig. 10(a)–(e), we plot the range cut for three methods at $r=23, 24, 25$ and Doppler frequency cut at $v=-0.20$ and $-0.18$. As can be seen from these firgues, all three algorithms still produce reasonable responses. In most cases, ADPM+RTR  still has the deepest nulls at scatterers locations.
\begin{figure}[htpb]
\centering
\subfigure[]
{
\centering
\includegraphics[width=1.6in]{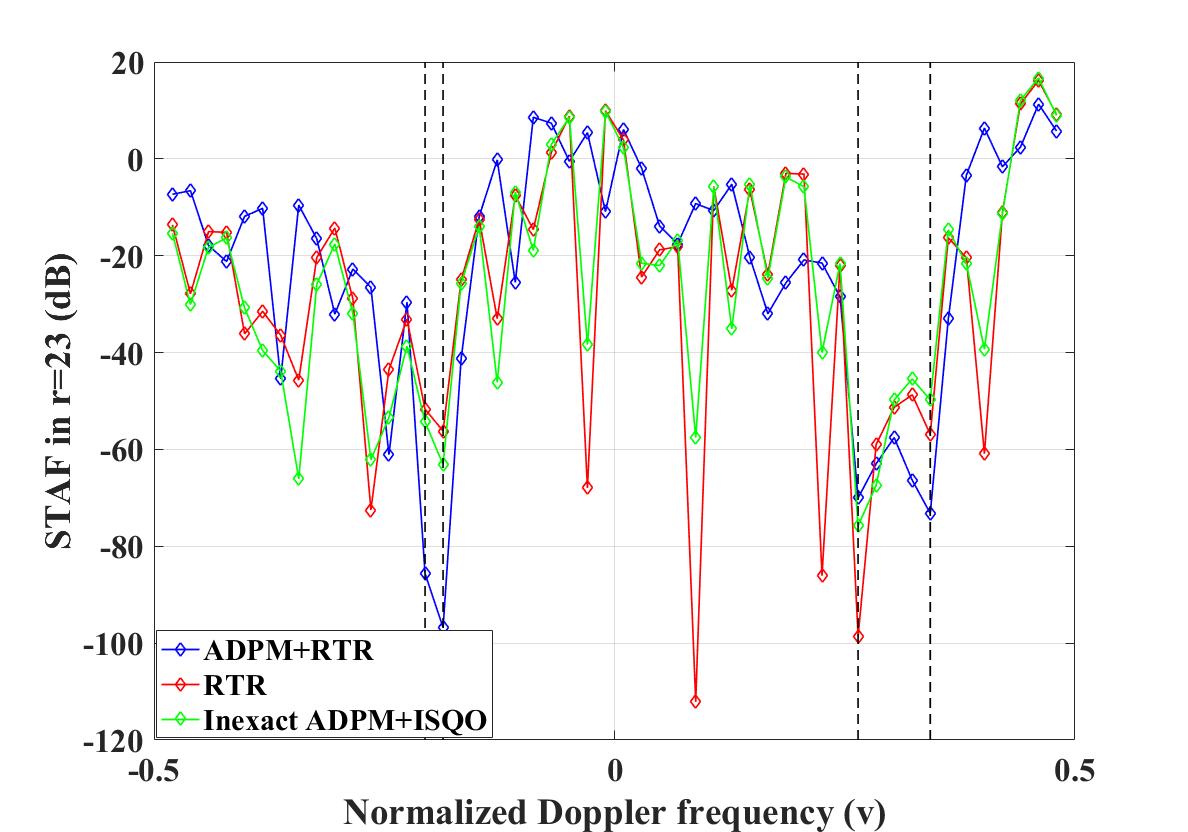}
\quad
}
\subfigure[]{
\centering
\includegraphics[width=1.6in]{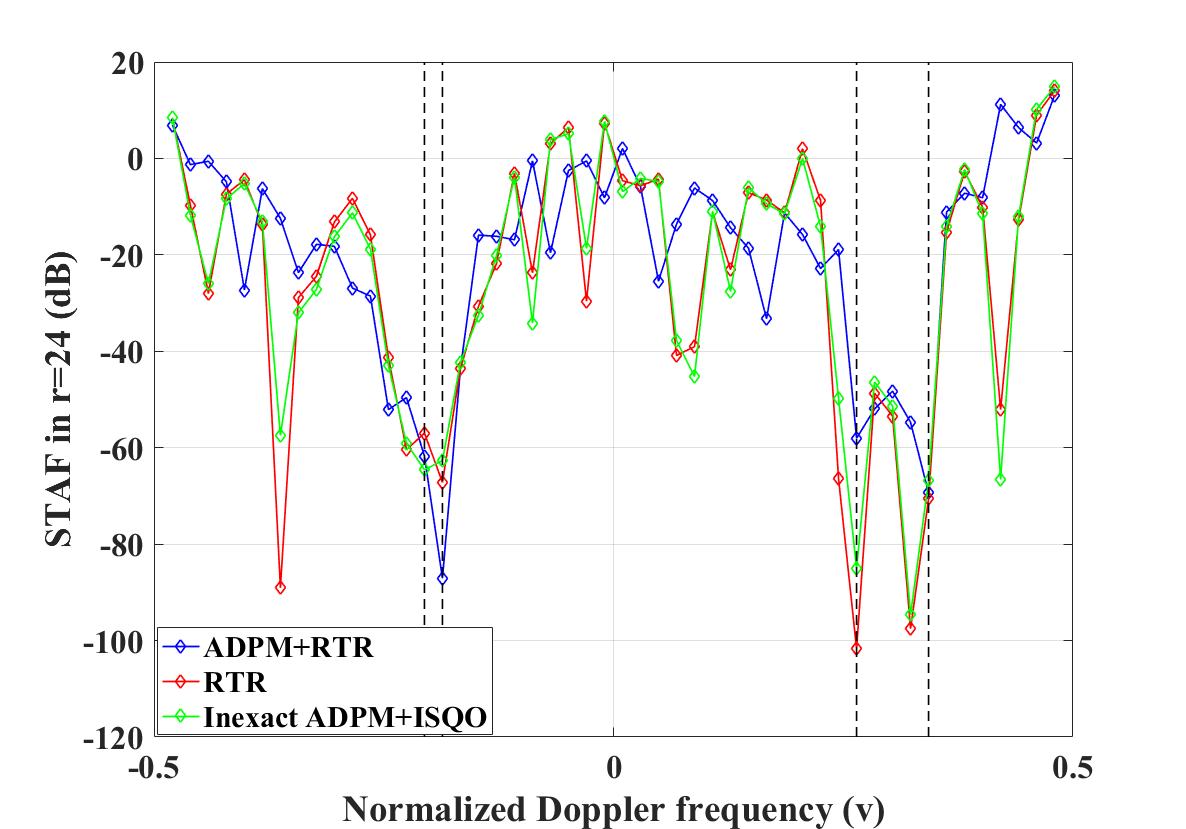}
\quad
}\\
\subfigure[]{
\centering
\includegraphics[width=1.6in]{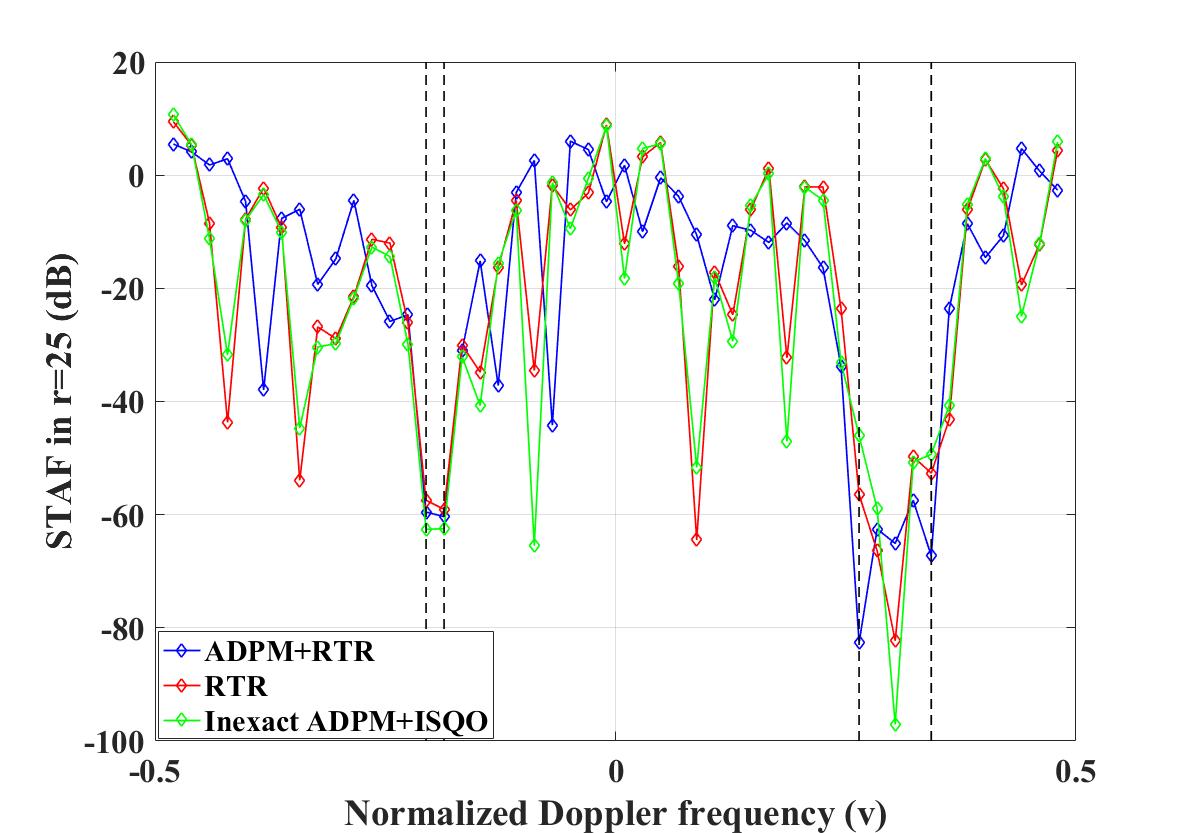}
\quad
}%
\subfigure[]{
\centering
\includegraphics[width=1.6in]{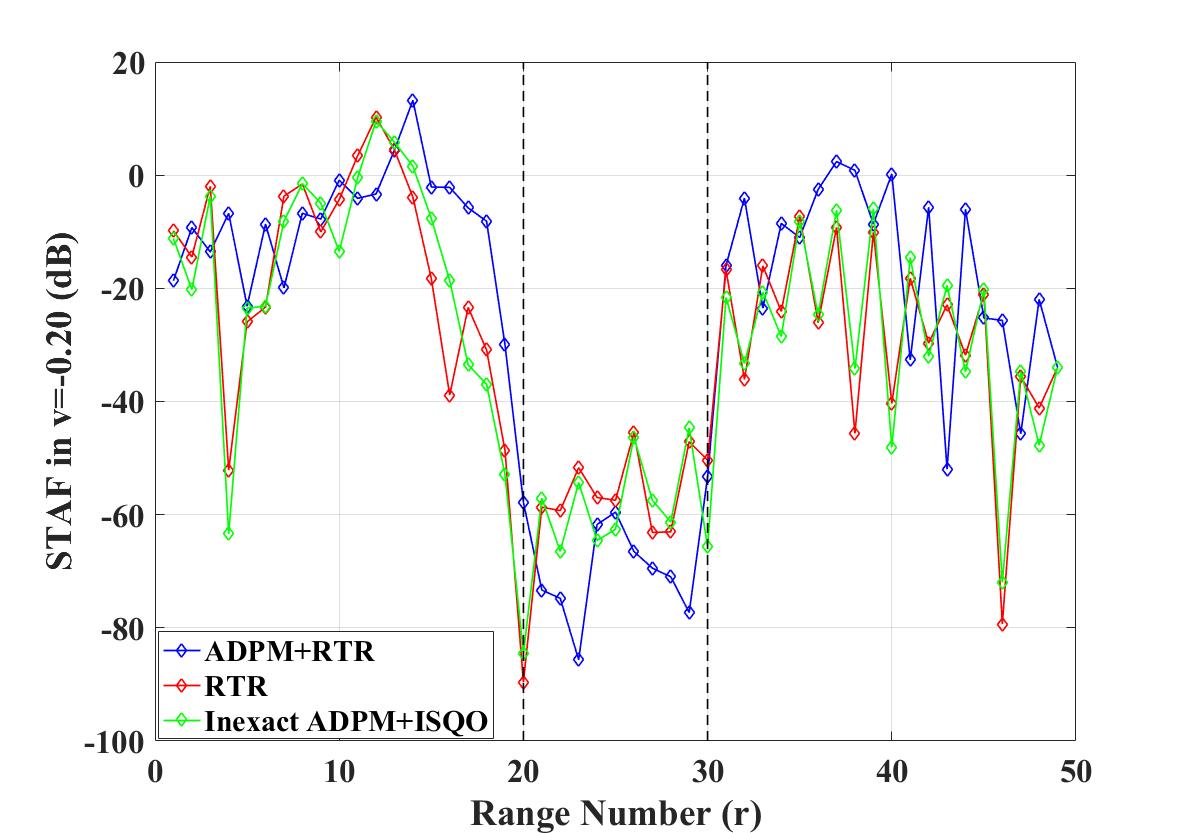}
\quad
}\\
\subfigure[]{
\centering
\includegraphics[width=1.6in]{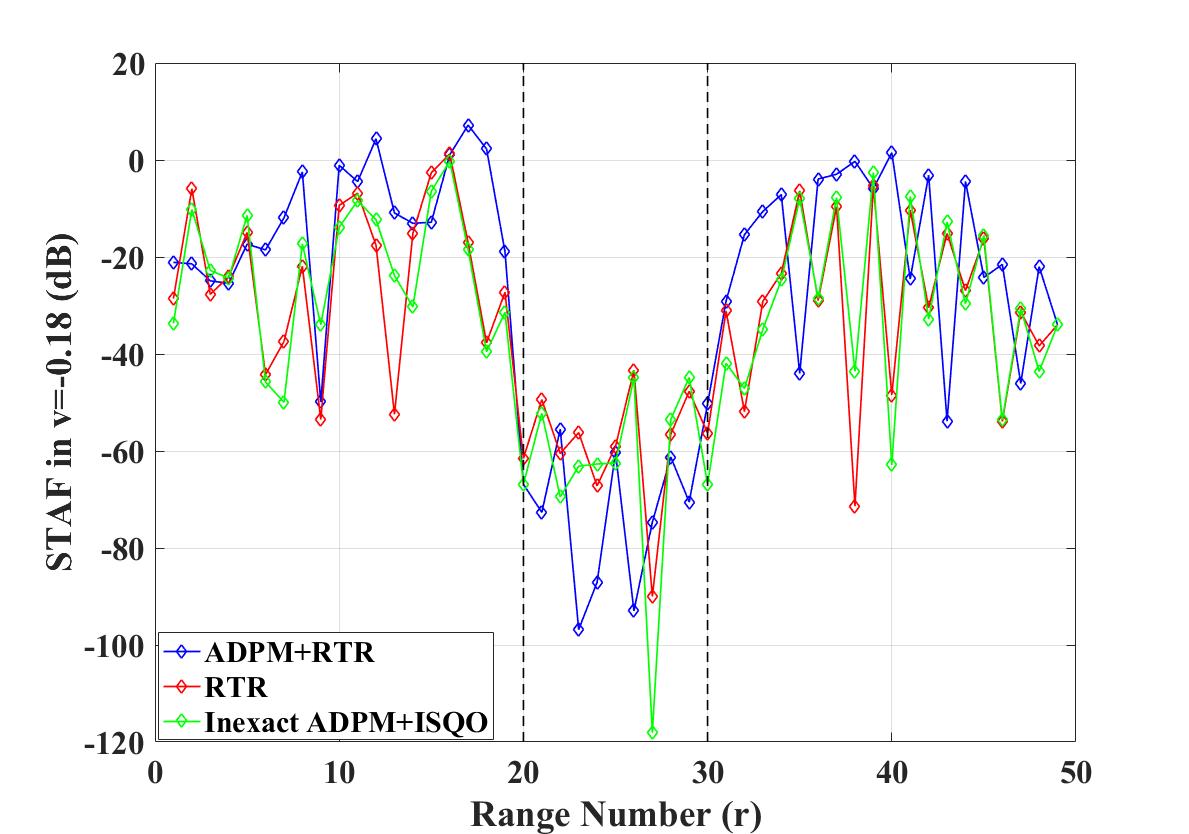}
\quad
} 
\caption{STAF cuts at (a) r = 23, (b) r=24, (c) r=25, (d) v=-0.20 and v=-0.18}
\label{3}
\end{figure}
The SIR average values and average simulation times in 2 of all three methods are plotted in Fig. 11. The ADPM+RTR algorithm still achieves the highest SIR values among three algorithms. Analysing the time consumption, the ADPM+RTR algorithm still takes more time for two-layer iterations.
\begin{figure}[htpb]
\centering
\centering
\includegraphics[width=3.2in]{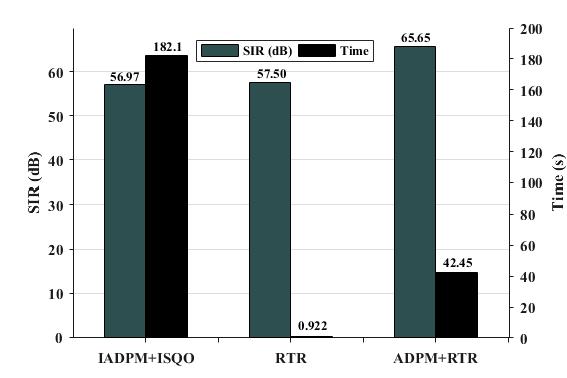}
\caption{SIR average values and average simulation times for Inexact ADPM+ISQO, RTR and ADPM+RTR.}
\end{figure}

\section{Conclusion}
In this paper, we introduced STAF shaping for CR under CMC. In order to solve the optimization problem with the non-convex polynomial under the CMC condition, an alternating direction Riemannian optimal algorithm is developed to minimize the cost function and thus further to improve the SIR value. In each iteration, it converts the non-convex multi-dimensional optimization problem into two sub-problems. One of the sub-problems is solved globally and the other sub-problem is utilized by complex circle Riemann manifold framework. Compared to previous algorithms, our proposed method gain higher SIR value. Combining the RTR algorithm, the internal iteration operation time is better than other conventional algorithms. In addition, the radar sequence optimized by this method can form a lower region of response values based on the interference distribution, with better interference suppression performance. Possible future research directions could involve the design of multi-input and multi-output radar waveform.

\begin{appendices}
\section{Appendix A}
According to (32), the directional derivative of the objective function $h({\mathbf{s}})$ is
\begin{align}\label{eq41}
Dh({\bf{s}})\left[ {{{\bf{\varsigma }}_{\bf{s}}}} \right] &= \frac{1}{a}\mathop {\lim }\limits_{a \to 0} [\mathop \sum \limits_{i = 0}^{K{N_v} - 1} {({\bf{s}} + a{{\bf{\varsigma }}_{\bf{s}}})^H}{{\bf{\Phi }}_i}({\bf{s}} + a{{\bf{\varsigma }}_{\bf{s}}})\\\nonumber
&{({\bf{s}} + a{{\bf{\varsigma }}_{\bf{s}}})^H}{\bf{\Phi }}_i^H({\bf{s}} + a{{\bf{\varsigma }}_{\bf{s}}})\\\nonumber
 &+ \frac{1}{2}{\rho ^{(r - 1)}}({\bf{s}} + a{{\bf{\varsigma }}_{\bf{s}}} - {{\bf{v}}^{(r)}}){({\bf{s}} + a{{\bf{\varsigma }}_{\bf{s}}} - {{\bf{v}}^{(r)}})^H}]\\\nonumber
 &- \frac{1}{a}\mathop {\lim }\limits_{a \to 0} [\sum\limits_{i = 0}^{K{N_v} - 1} {{{\bf{s}}^H}} {{\bf{\Phi }}_i}{\bf{s}}{{\bf{s}}^H}{\bf{\Phi }}_i^H{\bf{s}}\\\nonumber
 &+ \frac{1}{2}{\rho ^{(r - 1)}}({\bf{s}} - {{\bf{v}}^{(r)}}){({\bf{s}} - {{\bf{v}}^{(r)}})^H}]\\\nonumber
 &= \frac{1}{a}\mathop {\lim }\limits_{a \to 0} [\sum\limits_{i = 0}^{K{N_v} - 1} {({{\bf{s}}^H}{{\bf{\Phi }}_i}a{{\bf{\varsigma }}_{\bf{s}}}{{\bf{s}}^H}{\bf{\Phi }}_i^H{\bf{s}}}\\\nonumber  
 &+ a{\bf{\varsigma }}_{\bf{s}}^H{{\bf{\Phi }}_i}{\bf{s}}{{\bf{s}}^H}{\bf{\Phi }}_i^H{\bf{s}}+ {{\bf{s}}^H}{{\bf{\Phi }}_i}{\bf{s}}a{{\bf{\varsigma }}_{\bf{s}}}^H{\bf{\Phi }}_i^H{\bf{s}}\\\nonumber
 &+ {{\bf{s}}^H}{{\bf{\Phi }}_i}{\bf{s}}{{\bf{s}}^H}{\bf{\Phi }}_i^Ha{{\bf{\varsigma }}_{\bf{s}}})- a{\rho ^{(r - 1)}}({\bf{s}} - {{\bf{v}}^{(r)}}){{\bf{\varsigma }}_{\bf{s}}}^H]
\end{align}
The directional derivative of the Euclidean gradient Gradh(s) can be calculated by definition as
\begin{align}\label{eq42}
\begin{array}{*{20}{lll}}
{\rm{Grad}}h({\bf{s}})&=2\sum\limits_{i = 0}^{K{N_v}-1} {{{\bf{s}}^H}{\Phi _i}{\bf{s}}\Phi _i^H{\bf{s}} + {{\bf{s}}^H}\Phi _i^H{\bf{s}}{\Phi _i}{\bf{s}}}\\ &+ {\rho ^{(r - 1)}}({\bf{s}} - {\bf{v}}^{({r})}).
\end{array}
\end{align}
Combining equation (33) and (42), we have
\begin{align}\label{eq43}
{\rm{grad }}h({\bf{s}}) &= 2(\sum\limits_{i = 0}^{K{N_v} - 1} {{{\bf{s}}^H}{\Phi _i}{\bf{s}}\Phi _i^H{\bf{s}} + {{\bf{s}}^H}\Phi _i^H{\bf{s}}{\Phi _i}{\bf{s}}} )\\\nonumber
 &+ {\rho ^{(r - 1)}}({\bf{s}} - {{\bf{v}}^{(r)}})\\\nonumber
 &- 2{\mathcal R}\{ (\sum\limits_{i = 0}^{K{N_v} - 1} {{{\bf{s}}^H}{\Phi _i}{\bf{s}}\Phi _i^H{\bf{s}} + {{\bf{s}}^H}\Phi _i^H{\bf{s}}{\Phi _i}{\bf{s}})} \\\nonumber
 &+ {\rho ^{(r - 1)}}({\bf{s}} - {{\bf{v}}^{(r)}}) \odot {{\bf{s}}^*}\}  \odot {\bf{s}}
\end{align}
Similarly, directional derivative of Euclidean gradient ${\rm{DGrad}}h({\bf{s}})$ can be calculated as
\begin{align}\label{eq44}
{\rm{DGrad}}h({\bf{s}})[{{\bf{\varsigma }}_{\bf{s}}}] &= \frac{{\mathop {\lim }\limits_{a \to 0} [{\rm{Grad}}h({\bf{s}} + a{{\bf{\varsigma }}_{\bf{s}}}) - {\rm{Grad}}h({\bf{s}})]}}{a}\\\nonumber
 &= \frac{2}{a}\mathop {\lim }\limits_{a \to 0} [\mathop \sum \limits_{i = 0}^{K{N_v} - 1} a{\bf{\varsigma }}_{\bf{s}}^H{{\bf{\Phi }}_i}{\bf{s\Phi }}_i^H{\bf{s}}\\\nonumber
 &+ {{\bf{s}}^H}{{\bf{\Phi }}_i}a{{\bf{\varsigma }}_{\bf{s}}}{\bf{\Phi }}_i^H{\bf{s}} + {{\bf{s}}^H}{{\bf{\Phi }}_i}{\bf{s\Phi }}_i^Ha{{\bf{\varsigma }}_{\bf{s}}}\\\nonumber
 &+ {{\bf{\Phi }}_i}a{{\bf{\varsigma }}_{\bf{s}}}({{\bf{s}}^H}{\bf{\Phi }}_i^H{\bf{s}}) + {{\bf{\Phi }}_i}{\bf{s}}(a{\bf{\varsigma }}_{\bf{s}}^H{\bf{\Phi }}_i^H{\bf{s}})\\\nonumber
 &+ {{\bf{\Phi }}_i}{\bf{s}}({{\bf{s}}^H}{\bf{\Phi }}_i^Ha{{\bf{\varsigma }}_{\bf{s}}}) + \frac{1}{2}{\rho ^{(r - 1)}}a{{\bf{\varsigma }}_{\bf{s}}}]\\\nonumber
 &= 2[\mathop \sum \limits_{i = 0}^{K{N_v} - 1} {\bf{\varsigma }}_{\bf{s}}^H{{\bf{\Phi }}_i}{\bf{s\Phi }}_i^H{\bf{s}} + {{\bf{s}}^H}{{\bf{\Phi }}_i}{{\bf{\varsigma }}_{\bf{s}}}{\bf{\Phi }}_i^H{\bf{s}}\\\nonumber
 &+ {{\bf{s}}^H}{{\bf{\Phi }}_i}{\bf{s\Phi }}_i^H{{\bf{\varsigma }}_{\bf{s}}} + {{\bf{\Phi }}_i}{{\bf{\varsigma }}_{\bf{s}}}({{\bf{s}}^H}{\bf{\Phi }}_i^H{\bf{s}})\\\nonumber
 &+ {{\bf{\Phi }}_i}{\bf{s}}({\bf{\varsigma }}_{\bf{s}}^H{\bf{\Phi }}_i^H{\bf{s}})+\frac{1}{2}{\rho ^{(r - 1)}}{{\bf{\varsigma }}_{\bf{s}}}]
\end{align}
The directional derivative of the Euclidean Hessian matrix Hessh(s) can be calculated by definition as
\begin{align}\label{eq45}
{\rm{Hess }}h({\bf{s}}) &= 2[\sum\limits_{i = 0}^{K{N_v} - 1} {{{({{\bf{s}}^H}\Phi _i^H{\bf{s}}{\Phi _i})}^H} + {{\bf{s}}^H}\Phi _i^H{\bf{s}}{\Phi _i}} \\\nonumber
 &+ \Phi _i^H{\bf{s}}{({\Phi _i}{\bf{s}})^H} + \Phi _i^H{\bf{ss}}{\Phi _i} + {\Phi _i}{\bf{s}}{{\bf{s}}^H}{\Phi _i}\\\nonumber
 &+ ({\Phi _i}{\bf{s}}){({\Phi _i}{\bf{s}})^H}] + {\rho ^{(r - 1)}}{{\bf{I}}_K}.
\end{align}
Combining equation (33) and (45), the Riemainnian Hessian matrix is
\begin{align}\label{eq46}
{\rm{hess }}h({\bf{s}}) &= 2[\sum\limits_{i = 0}^{K{N_v} - 1} {{{({{\bf{s}}^H}\Phi _i^H{\bf{s}}{\Phi _i})}^H} + {{\bf{s}}^H}\Phi _i^H{\bf{s}}{\Phi _i}} \\\nonumber
 &+ \Phi _i^H{\bf{s}}{({\Phi _i}{\bf{s}})^H} + \Phi _i^H{\bf{ss}}{\Phi _i} + {\Phi _i}{\bf{s}}{{\bf{s}}^H}{\Phi _i}\\\nonumber
 &+ ({\Phi _i}{\bf{s}}){({\Phi _i}{\bf{s}})^H}] + {\rho ^{(r - 1)}}{{\bf{I}}_K}\\\nonumber
 &- 2{\mathcal R}\{ \{ [\sum\limits_{i = 0}^{K{N_v} - 1} {{{({{\bf{s}}^H}\Phi _i^H{\bf{s}}{\Phi _i})}^H} + {{\bf{s}}^H}\Phi _i^H{\bf{s}}{\Phi _i}} \\\nonumber
 &+ \Phi _i^H{\bf{s}}{({\Phi _i}{\bf{s}})^H} + \Phi _i^H{\bf{ss}}{\Phi _i} + {\Phi _i}{\bf{s}}{{\bf{s}}^H}{\Phi _i}\\\nonumber
 &+ ({\Phi _i}{\bf{s}}){({\Phi _i}{\bf{s}})^H}] + \frac{1}{2}{\rho ^{(r - 1)}}{{\bf{I}}_K}\}  \odot {{\bf{s}}^*}\}  \odot {\bf{s}}
\end{align}

As such, we can solve (38), (39) and (40) until meeting the termination condition in Algorithm 2.
\end{appendices}
\end{document}